\documentstyle[12pt,a4]{article}
\makeatletter
 
 \@addtoreset{equation}{section}
\makeatother
\def\arg{\hbox{arg}}
\begin{document}
\title{Gravitational Lensing
           and Catastrophe Theory}
\author{Yuuko KAKIGI, Takashi OKAMURA\thanks{Present Address :
          Department of Physics, Tokyo Institute of Technology,
          Oh-Okayama, Megro-ku, Tokyo, 152 Japan}~
       and Takeshi FUKUYAMA \\
        Department of Physics, \\
        Ritsumeikan University, Kusatsu, \\
        Shiga, 525 Japan}
\date{}
\maketitle
\begin{abstract}
Singularities of caustics appeared in gravitational lensing effect are
discussed analytically.
\par
Multipole expansion model of lensing object is mainly studied since it is
tractable and universal.
Our analyses are confirmed by numerical calculations and applied to multiple
quasar system of PG1115+080.
Consistencies with elliptical lens models are also discussed.
\end{abstract}

\vskip 7cm
PACS number  95.30.Sf, 98.80.Es
\vfill\eject
\section{Introduction}
\par
Our starting point in discussing gravitational lensing
(hereafter GL) is Fermat's potential \cite{SEF}, \cite{BK}.
Lens equation is given by the extremum of Fermat's potential
and the critical lines are points where the Jacobian
from source plane to lens plane vanishes. (See Fig.1)
\par
Fermat's potential is all from which we get information.
Assuming thin lens approximation, Fermat's potential
is given by Eq.(\ref{vecfermat}).
Namely, Fermat's potential $\phi$ is fixed when we have determined
the surface mass density $\Sigma (\vec x)$ normalized by $\Sigma_{crit}$~.
\par
$\phi$ depends on the cosmological large scale structure through
$\Sigma_{crit}$ whereas $\Sigma (\vec x)$ does on the lens models.
The former problem is, for instance, concerned with the information of
inhomogeneity of
the universe, Dyer-Roeder distance, cosmological parameters, etc.
The latter is related with the information of lensing object.
\par
Our arguments in this article are essentially restricted in the latter problem
----- model problem.
\par
As models, we may consider a variety of elliptical lens models that have
different power dependence of distance in mass distribution.
However their analytical surveys are not tractable and dedicated to numerical
calculations.
On the other hand, multipole expansion model seems to be rather tractable.
\par
In any case, the more models are sophisticated, the more parameter
space gets complicated.
Parameters are, for instance, mass, finite core size, ellipticity,
multipole moments and their angles, etc.
Such parameters are fixed to reproduce the observed quantities
such as image positions, image amplifications
and time delays between several images.
However such parameter fitting becomes ambiguous and
best fitting may not be unique in models with many parameters.
Even if parameters had been fitted uniquely, it does not mean
systematic understanding of GL.
\par
More concretely speaking, observed data are biased in many cases by complicated
unknown reasons. For instance, observed image amplifications are disturbed by
the dust lying between the source and observer. Also the magnitude of such
disturbance may be different in the respective light path. In that case it is
not
necessarily reliable that we fit the parameters naively to the raw observed
data. These problems are rather common to the observed cosmology. However they
are
especially serious in GL, since we have no confirmed physical meaning in mass
distribution model of lensing objects.
\par
So we need some strategy for searching the universal structure
of GL that is irrelevant to models and boundary conditions.
This is another problem --- formulation problem.
We apply catastrophe theory(CT) \cite{Arnold} for such strategy.
\par
The application of CT to GL has already been performed in gravitational
microlensing and proved to be very useful \cite{CR}.
The concepts of optical depth, the time duration and amplification
of image are made clear in the framework of CT.
\par
This paper is organized as follows.
In \S 2 we argue general framework of singularities of caustics
in complex coordinates, which is applied
to the concrete models in subsequent sections.
In \S 3 we discuss multipole expansion model.
We will see how CT work in gravitational lensing.
However it is indispensable to consider the transition of singularities
analytically for that purpose.
These analyses are confirmed by numerical calculations in \S 4 and applied to
the real multiple quasar system.
In the last section we consider elliptical lens model and see the consistency
with the multipole expansion model.

\vfill\eject
\section{General framework in complex coordinates}
\par
Assuming the thin lens approximation, Fermat's potential
is given by

\begin{equation}
 \phi (\vec x,\vec y)={1 \over 2} (\vec x-\vec y)^2
   - {1 \over \pi} \int d^2x' \kappa(\vec x')\ln |\vec x-\vec x'|~.
\label{vecfermat}
\end{equation}

Here $\vec x$ and $\vec y$ are two dimensional coordinates
of image and source positions,
respectively, which are illustrated in Fig.1~.

\begin{center}
  ----------  \\
     Fig.1    \\
  ----------
\end{center}

$\kappa$ is the surface mass density of lens object,

\begin{equation}
 \kappa(\vec x) =\Sigma(\vec x)/\Sigma_{crit}~,
\end{equation}

normalized by

$$ \Sigma_{crit} \equiv {D_{OS} \over {4\pi G D_{OL} D_{LS}}}~. $$

Thus we deal with the mapping of lens plane ${\bf R^2}$
 to source plane ${\bf R^2}$~.
\par
So it becomes very convenient to introduce the complex coordinates,

\begin{equation}
 (u_1,u_2)\equiv(u,\bar u)\equiv(x_1+i~x_2~,~x_1-i~x_2)~.
\end{equation}

Hereafter, we use Latin letters as the indices for real coordinates and
Greek ones as the indices for complex coordinates.
And we use the notation for derivatives,
 $\phi_i = \partial_i \phi=\partial \phi/ \partial x_i$, etc.
Then we obtain the following expressions for the various quantities
used to classify the singularities of caustics \cite{SEF},

\begin{equation}
 D \equiv \det(\phi_{ij}) \equiv -4~\tilde D = -4 \det(\phi_{\alpha\beta})
   = 4(|\phi_{u\bar u}|^2-|\phi_{uu}|^2~) = (1-\kappa)^2 - |\gamma|^2~,
\label{jacobian}
\end{equation}

\begin{equation}
 \Delta \phi \equiv \hbox{Tr}(\phi_{ij})=4 \phi_{u \bar u}=2(1-\kappa)~,
\label{poisson}
\end{equation}

\begin{equation}
 \gamma=2 \phi_{\bar u\bar u}~,
\label{shear}
\end{equation}

\begin{equation}
 \Delta \equiv \det(D_{ij})\equiv 16 \tilde\Delta
        =16 \det(\tilde D_{\alpha\beta})~,
\label{laplacian}
\end{equation}
where $\gamma$ is the shear.
Critical line is a curve in image plane characterized by $D=0$
where the amplification of images $\mu \equiv {1 \over D}$ becomes $\infty$ .
Next the quantity that we often use in the classification of
the singularities is the derivative operator along the critical line.

\begin{equation}
 (\vec T\cdot \nabla)=T_u \partial_ u + T_{\bar u} \partial_{\bar u}~,
\end{equation}

where $T_u$ is a tangential vector to critical line,

\begin{equation}
 T_u \equiv {1\over 2~i}\partial_{\bar u}\tilde D~.
\label{tantocrit}
\end{equation}

So we obtain the expression

\begin{equation}
 \hat L \equiv (\vec T\cdot \nabla)={1\over 2~i}
 (\partial_{\bar u}\tilde D\partial_u - \partial_u\tilde D\partial_{\bar u})~.
\end{equation}

\par
Here we can represent the conditions for various singularities
in the complex coordinates.

\subsection{Singularities of Caustics}
\par
\begin{description}
\item[1.] Cuspoid sequence
  \begin{description}
  \item[1-a.] Fold singularity :

    \begin{equation}
  	 0=4\tilde D=-[(1-\kappa)^2-|2 \phi_{uu}|^2]
    \label{fold}
    \end{equation}

  \item[1-b.] Cusp singularity : Besides 1-a condition,

    \begin{equation}
	   0=\hat L \phi_u={1\over 2~i}
	   (\partial_{\bar u}\tilde D~\phi_{uu}
	   - \partial_u\tilde D~\phi_{u \bar u})
    \label{cusp}
    \end{equation}

  \item[1-c.] Swallowtail singularity : Besides 1-b conditions,

    \begin{equation}
	   0=\hat L^2 \phi_u
    \label{swallowtail}
    \end{equation}

  \item[1-d.] Butterfly singularity : Besides 1-c conditions,

    \begin{equation}
	   0=\hat L^3 \phi_u
    \label{butterfly}
    \end{equation}

  \end{description}
\item[2.] Umbilic :

$$ \phi_{\alpha\beta}=0~~~~~~  (\alpha, \beta = u, \bar u) $$

  \begin{description}
  \item[2-a.] Elliptic Umbilic

    \begin{equation}
    \tilde\Delta > 0
    \label{elliptic}
    \end{equation}

  \item[2-b.] Hyperbolic Umbilic

    \begin{equation}
     \tilde\Delta < 0
    \label{hyperbolic}
    \end{equation}

  \item[2-c.] Parabolic Umbilic

    \begin{equation}
     \tilde\Delta = 0
    \label{parabolic}
    \end{equation}

  \end{description}

\item[3.] Beak-to-beak and Lips

$$ \partial_u \tilde D=0 $$

  \begin{description}

  \item[3-a.] Lips

    \begin{equation}
     \tilde\Delta > 0
    \label{lips}
    \end{equation}

  \item[3-b.] Beak-to-Beak

    \begin{equation}
     \tilde\Delta < 0
    \label{beaktobeak}
    \end{equation}

  \end{description}
\end{description}

\subsection{Shapes of images}
\par
Fermat's potential is given by

\begin{equation}
 \phi = {1 \over 2} |u-v|^2 - \psi(u, \bar u)~,
\label{cfermat}
\end{equation}

where $\psi$ is deflection potential and $v \equiv y_1 + i~y_2$ represents
source position.
Lens equation (Fermat's principle) gives

\begin{equation}
 v = u - 2\partial_{\bar u} \psi \equiv F (u, \bar u)~.
\label{clenseq}
\end{equation}

Differentiating Eq.(\ref{clenseq}), we obtain

\begin{eqnarray}
 du &=& {1 \over |F_u|^2 - |F_{\bar u}|^2}
 (\bar{F_u} dv - F_{\bar u} d\bar v)
 ={1\over D}[(1-\kappa)dv-\gamma d\bar v]~,
 \nonumber \\
 d\bar u &=& {1 \over |F_u|^2 - |F_{\bar u}|^2}
 (-\bar{F_{\bar u}} dv + F_u d\bar v)
 ={1\over D}[-\bar\gamma dv+(1-\kappa)d\bar v]~,
\label{displace}
\end{eqnarray}

where

$$ F_u \equiv 1 - 2\partial_u \partial_{\bar u} \psi = 1 - \kappa~~~~
	(\hbox{real})~, $$

and

\begin{equation}
 F_{\bar u} \equiv -2\partial_{\bar u}^2 \psi= \gamma~.
\end{equation}

\par
Therefore Eq.(\ref{jacobian}) is rewritten as

\begin{equation}
 D = |F_u|^2 - |F_{\bar u}|^2~.
\end{equation}

So $|du|^2$ becomes

\begin{equation}
 |du|^2 = {(1-\kappa)^2 |dv|^2 \over D^2}
  [1+|\lambda|^2-(\lambda {dv \over d\bar v}+c.c.)]~.
\label{dusquare}
\end{equation}

Here we have introduced the notation

\begin{equation}
 \lambda \equiv \bar {F_{\bar u}} / F_u = {\bar\gamma \over 1-\kappa}~.
\end{equation}

{}From Eqs.(\ref{displace}) and (\ref{dusquare}), we know that the $dv$ which
gives
 the major axis of an image satisfies,

\begin{equation}
 \arg(\lambda {dv \over d\bar v}) = \pi ~,
\end{equation}

Therefore argument of $du$ along major axis is

\begin{eqnarray}
 \arg(du) & = & {1 \over 2}\arg(\gamma) + {1 \over 2}\arg(1-\kappa)
                +{\pi \over 2} \\
         & = & {1 \over 2}\arg(\gamma) + {\pi\over 2}\theta(1-\kappa)~,~~~~
                \; \; \; \; \; \; \; \; \; \; \; \; (\hbox{mod.}~~ \pi)
         ~, \nonumber
\label{elongation}
\end{eqnarray}

where $\theta$ is step function.
{}From Eq.(\ref{tantocrit}), the vector tangential to critical line is

\begin{eqnarray}
 T_u & = & {1 \over 2i}~\tilde D_{\bar u} \\
   & = & {1 \over 4i} [\kappa_u \psi_{\bar u \bar u}
                       + 2\psi_{\bar u \bar u \bar u} \psi_{uu}
                       +(1-\kappa)\kappa_{\bar u}] \nonumber
\end{eqnarray}

Finally, we obtain the relative correlation between  the direction of image
 elongation and critical curve,

\begin{equation}
 \arg({du \over T_u}) = {1 \over 2}\arg(\gamma)
                     +{\pi \over 2}\theta(\kappa-1)
                     -\arg(\tilde D_{\bar u})
\label{elongtocrit}
\end{equation}

The meaning of Eq.(\ref{elongtocrit}) is explained in the concrete models in
subsequent
sections. Here we show how it works in the most simple model,
spherical model.
\par
In this case, $\psi = \psi(|u|^2)$~and

\begin{equation}
 \gamma = -2 \partial_{\bar u}^2 \psi = -2 u^2 \psi'' ~~~~\hbox{etc.}~,
\end{equation}

where $\psi' = {d\psi \over d\omega}$ with $\omega \equiv |u|^2$~.
\par
So we obtain the equation,

\begin{equation}
 \arg(du) = \arg~u + {1 \over 2} \arg(\psi'')
           +{\pi \over 2}\theta(\kappa-1)~.
\end{equation}

If we assume that $\kappa$ is monotonically decreasing function of
$\omega$, we obtain the inequality,

\begin{eqnarray}
 \psi'' & = & {1 \over \omega} ({\kappa \over 2}-\psi') \nonumber \\
        & = & {1 \over 2\omega^2}[\omega \kappa
                                    - \int^\omega_0 d\omega \kappa] < 0~.
\end{eqnarray}

Therefore the images are elongated along

\[ \left<
  \begin{array}{ll}
    \hbox{circular direction for}~\kappa -1 < 0~.  \\
    \hbox{radial direction for}~\kappa -1 > 0 ~.
  \end{array}
\right. \]

This conclusion is irrelevant to the image position relative
to the critical lines.

\vfill\eject

\section{Multipole Expansion}
\par
Multipole expansion is the expansion of Eq.(\ref{vecfermat})
in powers of ${1 \over |\vec x|}$ and is given by

\begin{equation}
 \phi={1 \over 2} (\vec x-\vec y)^2 -{m \over 2}\ln |\vec x|^2
	+{\vec x\cdot\vec d \over |\vec x|^2}
	+{{}^t \vec x \hat q \vec x \over |\vec x|^4}~,
\end{equation}

where $m$, $\vec d$, $\hat q$ are reduced total mass, dipole and
quadrupole moments of lens object, respectively.
The "codimension" is identical with the number of coordinates independent
equalities. And the condition whether the singularity is generic
or not is determined by
Dimension of parameter space $\ge$ Codimension.
State variables $\vec x$ and control parameters $\vec y, m, \vec d, \hat q$
are expressed in complex coordinates as

\begin{eqnarray}
 & z=x_1+ i x_2~, &~~~s=y_1+ i y_2~, \nonumber \\
 & d=d_1+ i d_2~, &~~~q=q_{11} + i q_{12}~~~~(q_{11}+q_{22}=0)~.
\label{def}
\end{eqnarray}

Taking, furthermore, the following rescaling,

\begin{equation}
 u={z \over m^{1 \over 2}}~,~~~v={s \over m^{1 \over 2}}~,~~~
 \delta={d \over m^{3\over2}}~,~~~Q={q \over m^2}~,
\end{equation}

we get Fermat's potential in terms of complex coordinates \cite{YO}.

\begin{equation}
 \bar\phi = {\phi \over m}= {1 \over 2} |u-v|^2-{1 \over 2}\ln |\bar u u|
	+{1 \over 2}({\delta \over u}+{\bar\delta \over \bar u})
	+{1 \over 2}({Q \over u^2}+{\bar Q \over \bar u^2})~.
\label{cmultifermat}
\end{equation}

The parameters in Eq.(\ref{cmultifermat}) are
associated to physical parameters as follows:

\begin{eqnarray}
 m & = & {M \over M_{crit}}~,~~~
             M_{crit}\equiv {D_{OL} D_{OS} \over 4 G D_{LS}}~,  \\
 \delta & = & {D_1+i D_2 \over M_{crit} D_{OL}}m^{-{3\over2}}~, \\
 Q & = & {I_{11}+i I_{12} \over M_{crit} D_{OL}^2}m^{-2}~,
\end{eqnarray}

where $M$, $D_\nu$ and $I_{\nu\lambda}$ correspond the physical
total mass, the physical 3-dimensional dipole moment and
the physical 3-dimensional quadrupole moment, respectively.
One of the aims of this article to find the relationships between
model parameters and singularities in analytical ways.
So we first discuss about the simplified model and the lowest
singularities.
Our task here is to express the equalities in \S 2 by the
multipole parameters.
\par
Critical point condition Eq.(\ref{fold}) becomes

\begin{equation}
 |u|^2 = |1+2{\delta \over u}+6{Q \over u^2}|~.
\label{multifold}
\end{equation}

Cusp condition Eq.(\ref{cusp}) is

\begin{equation}
 {(1+2{\delta \over u}+6{Q \over u^2})^3 \over
    (1+3{\delta \over u}+12{Q \over u^2})^2} =
  {|u|^6 \over |1+3{\delta \over u}+12{Q \over u^2}|^2}~.
\label{multicusp}
\end{equation}

Eqs.(\ref{multifold})and (\ref{multicusp}) are too complicated
for analytical surveys. So we discuss $\delta$ and $Q$ separately
in the following.

\subsection{Q = 0 CASE}
\par
At first, we consider Q = 0 case. Then Eqs.(\ref{multifold}) and
(\ref{multicusp}) are reduced to

\begin{equation}
 |u|^2 = |1+2{\delta \over u}|~,
\label{difold}
\end{equation}

\begin{equation}
 {(1+2{\delta \over u})^3 \over (1+3{\delta \over u})^2} =
  {|u|^6 \over |1+3{\delta \over u}|^2}~.
\label{dicusp}
\end{equation}

{}From Eq.(\ref{dicusp}), we obtain

\begin{equation}
 \hbox{Im}[{(1+2{\delta \over u})^3 \over (1+3{\delta \over u})^2}] = 0~,
\label{dicuspim}
\end{equation}

\begin{equation}
 {(1+2{\delta \over u})^3 \over (1+3{\delta \over u})^2} > 0~.
\label{dicuspre}
\end{equation}

We introduce the geometrical variables,

\begin{equation}
 {u \over \delta} \equiv {\tau \over |\delta|}e^{i \theta}
                  \equiv t e^{i \theta}~,
\end{equation}

where $\tau$ is real and $\theta$ is the physical angle of image position
 from the dipole direction.
In terms of the geometrical variables, Eqs.(\ref{difold}),
(\ref{dicuspim}) and (\ref{dicuspre})
are expressed by

\begin{equation}
 |\delta|^2 t^3 = \sqrt{t^2 + 4t\cos \theta + 4}~,
\label{difoldII}
\end{equation}

\begin{equation}
 \sin \theta [3t^3 \cos \theta + t^2 (5 + 16 \cos^2 \theta)
                 + 48t \cos \theta + 36] = 0~,
\label{dicuspimII}
\end{equation}

\begin{eqnarray}
 t^5 + 12t^4 \cos \theta +3t^3 (5 + 14 \cos^2 \theta)
          +2t^2 \cos \theta (51 + 16 \cos^2 \theta)
\nonumber \\
    + 12t (5 + 8 \cos^2 \theta) + 72 \cos \theta > 0~,
\label{dicuspreII}
\end{eqnarray}

respectively.

\vskip .5cm

(i) $\sin \theta = 0$
\par
At first, we consider the critical points on the axis of the dipole
direction.
Then Eq.(\ref{dicuspimII}) vanishes.
And Eqs.(\ref{difold}) and (\ref{dicuspreII}) are written as follows:

\begin{equation}
 |\delta|^4 y^6 = (y + 2)^2~,
\end{equation}

\begin{equation}
 y(y + 2) > 0~,
\label{sinzero}
\end{equation}

where $y \equiv t \cos \theta$. The solutions of Eq.(\ref{sinzero})
change at $|\delta| = \sqrt{3} / 9$ qualitatively.
At this value, beak-to-beak singularity
appears.
When $|\delta|$ is less than this value, cusp singularity appears.
And when $|\delta|$ is more than this value, fold singularity appears.
\vskip .5cm

(ii) $\sin \theta \ne 0$

\par
Off the axis, we regard Eqs.(\ref{difoldII})-(\ref{dicuspreII})
as the equations for $\delta, t \cos \theta (\equiv y)$ and
$t^2 (\equiv x)$.
Using $x$ and $y$, Eqs.(\ref{difoldII}) and (\ref{dicuspimII}) become

\begin{equation}
 |\delta|^4 x^3-x-4 = 4y
\label{difoldIII}
\end{equation}

and

\begin{equation}
 x=-{4(2y+3)^2 \over 3y+5}~~,
\label{dicuspimIII}
\end{equation}

respectively.
However, $x > 0, x \geq y^2$ by definition, which implies

$$ -3 \leq y < -{5 \over 3}~~. $$

Also Eq.(\ref{dicuspreII}) is rewritten as

\begin{equation}
 0 < {4x \over t}{(y+2)^3 (y+3) \over (3y+5)^2}~~.
\end{equation}

Consequently $y$ satisfies

\begin{equation}
-2 < y < -{5 \over 3}~.
\end{equation}

Eliminating $x$ from Eqs.(\ref{difoldIII}) and (\ref{dicuspimIII}),
 we obtain

\begin{equation}
 4|\delta|^2 (2y+3)^3=(3y+5)|y+2|~~.
\end{equation}

That is,

\begin{equation}
 |\delta|^2 = {9 \over 4}
             {\rho (1 - \rho) \over (2\rho + 1)^3} \equiv f(\rho)
\label{function}
\end{equation}

where $\rho$ is bounded

$$ 0 < \rho \equiv -(3t \cos \theta + 5) <1~. $$

The function $f(\rho)$ relation is plotted in Fig.2.

\begin{center}
  ----------  \\
     Fig.2    \\
  ----------
\end{center}

\par
Two solutions coalesce at $|\delta|^2 = {{10 + 7\sqrt{7}} \over 6^3}$ in which
 higher singularity, swallowtail, appears.
At $|\delta|$ more (less) than this value, cusp (fold) singularity appears.

\subsection{$\delta = 0$ CASE}
\par
Next, we consider $\delta = 0$ case. Then Eqs.(\ref{multifold}) and
(\ref{multicusp}) are

\begin{equation}
 |u|^2 = |1 + 6{Q \over u^2}|~,
\label{quafold}
\end{equation}

\begin{equation}
 {(1 + 6{Q \over u^2})^3 \over (1 + 12{Q \over u^2})^2} =
   {|u|^6 \over |1 + 12{Q \over u^2}|^2}~.
\label{quacusp}
\end{equation}

{}From Eq.(\ref{quacusp}) we obtain

\begin{equation}
 \hbox{Im}[{(1 + 6{Q \over u^2})^3 \over (1 + 12{Q \over u^2})^2}] = 0~,
\label{quacuspim}
\end{equation}

\begin{equation}
 {(1 + 6{Q \over u^2})^3 \over (1 + 12{Q \over u^2})^2} > 0~.
\label{quacuspre}
\end{equation}

We introduce the geometrical variables

\begin{equation}
 {u^2 \over Q} = {\tau^2 \over |Q|}e^{2i\theta} \equiv w e^{2i\theta}~,
\end{equation}

where $\tau$ is real and $\theta$ is the physical angle of image position from
the quadrupole direction. In terms of the geometrical variables,
Eqs.(\ref{quafold}),(\ref{quacuspim}) and (\ref{quacuspre})
are expressed by

\begin{equation}
 |Q|w^2 = \sqrt{w^2 + 12w \cos 2\theta +36}~,
\label{quafoldII}
\end{equation}

\begin{eqnarray}
 \sin 2\theta [w^4 +   12w^3 \cos 2\theta - 36 w^2
              (4\cos^2 2\theta - 1)
\nonumber \\
   -8 \cdot 6^3 w \cos 2\theta - 4 \cdot 6^4] = 0~,
\label{quacuspimII}
\end{eqnarray}

\begin{eqnarray}
 w^6 + 42w^5 \cos 2\theta  +   36w^4 (5 + 14 \cos^2 2\theta)
     + 6^3 w^3 \cos 2\theta (21 + 4 \cos^2 2\theta)
\nonumber \\
  +  8 \cdot 6^4 w^2 (1+\cos^2 2\theta) + 4 \cdot 6^5 w \cos 2\theta > 0~,
\label{quacuspreII}
\end{eqnarray}

respectively.

\vskip .5cm

(i) $\sin 2\theta = 0$
\par
At first, we consider the singularity on the axis of the quadrupole direction
and on the axis normal to it. Then Eqs.(\ref{quafoldII}) and
(\ref{quacuspreII}) are written as follows:
$$ |Q|^2 y^4 = (y+6)^2~, $$

\begin{equation}
 y(y+6)>0~,
\label{quacuspreIII}
\end{equation}

where $y \equiv |u/Q|^2 \cos 2\theta$. The solutions of
Eqs.(\ref{quacuspreIII}) change at
$|Q|=1/24$ qualitatively.
At this value, beak-to-beak singularity appears.
When $|Q|$ is less than this value, cusp singularity appears.
And when $|Q|$ is more than this value, fold singularity appears.

\vskip .5cm

(ii) $\sin 2\theta \ne 0$
\par
we regard Eqs.(\ref{quafoldII}) - (\ref{quacuspreII})
at points off those axes as the equations for $Q,
w \cos 2\theta$ and $w^2$. Those solutions of
Eqs.(\ref{quafoldII}) - (\ref{quacuspreII}) qualitatively
change at $|Q|={\sqrt{598+82\sqrt{41}} \over 192}$ and $|Q|=1/8$.
They correspond to swallowtail and butterfly singularity, respectively.
When $|Q|$ is less than the swallowtail-value, cusp singularity appears.
And when $|Q|$ is more than that value, fold singularity appears.
Talking about butterfly singularity, when $|Q|$ is not butterfly-value,
cusp singularity always appears.
These results are confirmed by numerical calculations.

\subsection{Shapes of images}
\par
{}From Eq.(\ref{tantocrit}) the vector tangential to critical line is

\begin{equation}
 T_u={1 \over 2i} \psi_{uu} \psi_{\bar u \bar u \bar u}
  ={1 \over 8i}\bar\gamma \partial_{\bar u}\gamma
\label{multantocrit}
\end{equation}

for non-transparent lens model like multipole expansion.
{}From Eqs.(\ref{multantocrit}) and (\ref{elongtocrit}) it follows

\begin{eqnarray}
 \arg({du \over T_u}) & = & {3 \over 2}\arg(\gamma)
 -\arg(\partial_{\bar u} \gamma)
                                                 \nonumber \\
   & = & {\pi \over 2}+{1 \over 2}\arg[{(\psi_{\bar u \bar u})^3
                  \over (\psi_{\bar u \bar u \bar u})^2}]
\end{eqnarray}
\par
If source position is in the neighborhood of cusp singularity,
${(\psi_{\bar u \bar u})^3 \over (\psi_{\bar u \bar u \bar u})^2}$ is
concluded to be almost negative definite from Eq.(\ref{multicusp}).
Therefore,

\begin{equation}
 \arg({du \over T_u}) \sim  0~~\hbox{or}~~\pi~.
\end{equation}

Namely, the images in the neighborhood of cusp are
parallel to critical line.
\par
For the general source position,

\begin{equation}
 \arg({du \over T_u})={3 \over 2}\arg[1+2{\bar \delta \over \bar u}
                                      +6{\bar Q \over \bar u^2}]
     -\arg[1+3{\bar \delta \over \bar u}+12{\bar Q \over \bar u^2}]~.
\end{equation}

\vfill\eject

\section{Numerical Calculations}
\par
In the previous section we have analyzed lens equation in which lensing
object is modeled by multipole expansion.
In this section we will check the results of each case by numerical
calculations in the first two subsections.
In the last subsection we will apply multipole expansion model to PG1115+080
\cite{WCW},
which is compared with the article by Yoshida and Omote \cite{YO}.

\subsection{Q = 0 case}
\par
Fig.3 shows the behaviors of caustics and critical lines
in the neighborhood of $|\delta| = {\sqrt{3} \over 9}$
at which beak-to-beak appears on axis.

\begin{center}
  ----------  \\
     Fig.3    \\
  ----------
\end{center}

In $|\delta| < {\sqrt{3} \over 9}$, the inequality
in Eqs.(\ref{sinzero}) is satisfied and cusp appears on axis (Fig.3a).
At $|\delta| = {\sqrt{3} \over 9}$ cusps coalesce to
beak-to-beak on axis (Fig.3b).
In  $|\delta| > {\sqrt{3} \over 9}$, the inequality in
Eqs.(\ref{sinzero}) is not satisfied. So cusps disappear
and fold is born (Fig.3c).
\par
Fig.4 shows the behaviors in the neighborhood of
$|\delta|^2 = {{10 + 7\sqrt{7}} \over 6^3}$
at which swallowtail appears off axis.

\begin{center}
  ----------  \\
     Fig.4    \\
  ----------
\end{center}

This corresponds to the argument in subsection \S 3.1(ii).
As was indicated there, cusp singuralities at points P change
to swallowtail when $|\delta|$ increases to
$\delta_c \equiv \sqrt{{{10 + 7\sqrt{7}} \over 6^3}}$
(Fig.4a and Fig.4b) and cusps disappear and fold appears(Fig.4c).
Swallowtail singularity is easily checked by Fig.4a corresponding to
$|\delta| < \delta_c$.

\subsection{$|\delta| = 0$ Case}
\par
First we simulate the case (i) in \S 3-2.
Fig.5 shows the behaviors of singularities in the neighborhood of
$|Q| = {1 \over 24}$ at which beak-to-beak appears on
$\theta = {\pi \over 2}$ axis.

\begin{center}
  ----------  \\
     Fig.5    \\
  ----------
\end{center}

At $|Q| < {1 \over 24}$, the inequality in Eqs.(\ref{quacuspreIII})
is satisfied and cusps appear (Fig.5a).
They coalesce to beak-to-beak at $|Q| = {1 \over 24}$ on the $Y$-axis
(Fig.5b). The inequality in Eqs.(\ref{quacuspreIII})
is not satisfied at $|Q| > {1 \over 24}$ and fold appears
instead of cusp (Fig.5c).
\par
Furthermore the case (ii) is simulated in Fig.6 and Fig.7.

\begin{center}
  ----------  \\
     Fig.6    \\
  ----------
\end{center}
\begin{center}
  ----------  \\
     Fig.7    \\
  ----------
\end{center}

$|Q|$ increases passing through two critical points,
${1 \over 8} (\equiv Q_1)$ and
${\sqrt{598+82\sqrt{41}} \over 192} (\equiv Q_2)$.
Fig.6b and Fig.7b correspond to the transition points of butterfly
and swallowtail, respectively.
This is confirmed from Fig.6c and Fig.7a that are diagrams
with $|Q| > Q_1$
and $|Q| < Q_2$, respectively.

\subsection{Application to multiple quasar}
\par
In this subsection we apply multipole expansion model
Eq.(\ref{cmultifermat}) to lensing system PG1115+080.
We adopt the observed data by Christian et al \cite{CCW}, which is
exhibited in Table 1.

\begin{center}
  -------------  \\
     Table 1     \\
  -------------
\end{center}

Numerical result is shown in Fig.8.

\begin{center}
  ----------  \\
     Fig.8    \\
  ----------
\end{center}

Here we have taken the values for control parameters in
Eq.(\ref{cmultifermat}) as follows:

$$ \hbox{source position} = (0^{\prime \prime}.009, 0^{\prime \prime}.021)~, $$

\begin{equation}
 m = 3.468 \times 10^{-11}~,
\end{equation}

$$ |\delta| = 0.1912~,~~~~~~~\theta_{\delta} = 91^\circ .85~,    $$

$$ |Q| = 0.0864~,~~~~~~~\theta_Q = 135^\circ .79~,     $$

where $\theta_{\delta}$ and $\theta_Q$ are arguments of $\delta$ and $Q$,
respectively.
The parameters adopted in \cite{YO} are related to our parameters
by
\begin{eqnarray}
 \mu &=& m~ \hbox{(mass)}~, \nonumber \\
 |\delta| &=& m^{1/2}|\delta_{our}| = 1.126\times 10^{-6}~, \nonumber \\
 |\xi| &=& 2 m |Q| = 5.993 \times 10^{-12}~, \nonumber \\
 \chi_d &=& \theta_\delta - {\theta_Q \over 2} = 23^\circ .95~, \nonumber \\
 \chi_G &=& {\theta_Q \over 2} = 67^\circ .90~.  \nonumber \\
\end{eqnarray}

\par
Calculated image positions are

\begin{eqnarray}
 \theta(A_1) &=& (-0^{\prime \prime}.84, -0^{\prime \prime}.87)~, \nonumber \\
 \theta(A_2) &=& (-1^{\prime \prime}.12, -0^{\prime \prime}.26)~, \nonumber \\
 \theta(B) &=& (0^{\prime \prime}.70, -0^{\prime \prime}.60)~, \\
 \theta(C) &=& (0^{\prime \prime}.35, 1^{\prime \prime}.34)~, \nonumber \\
 \theta(D) &=& (0^{\prime \prime}.56, -0^{\prime \prime}.42)~, \nonumber \\
 \theta(E) &=& (-0^{\prime \prime}.49, 0^{\prime \prime}.08)~, \nonumber
\end{eqnarray}

where we had the additional two images labeled by D and E as was
indicated by Ref.\cite{YO}.
Here we show source position and image positions $\theta$ in unnormalized
values that can be compared with observed values,
while we use variables $u$, $v$ normalized by $\sqrt m$ in
Eq.(\ref{cmultifermat}).

\par
The relative magnifications of images are also calculated from
Eq.(\ref{jacobian}) as

\begin{eqnarray}
 |{\mu_{A_1} \over \mu_C}| &=& 2.55~,~~~
 |{\mu_{A_2} \over \mu_C}|=2.22~,~~~
 |{\mu_B \over \mu_C}|=0.78~,  \\
 |{\mu_D \over \mu_C}| &=& 0.15~,~~~~
 |{\mu_E \over \mu_C}|=0.73 \times 10^{-2}~. \nonumber
\end{eqnarray}

We will check the validity of approximation in multipole expansion.
It must hold the following conditions

\begin{equation}
 1 \gg {|\delta| \over |u|} \gg {|Q| \over |u|^2}
\label{validity}
\end{equation}

The best fit parameters give

\begin{equation}
 {|\delta| \over |u|} \approx 0.2266  ~~~\hbox{and}~~
 {|Q| \over |u|^2} \approx 0.1390
\end{equation}

Thus the condition Eq.(\ref{validity}) is rather marginal.
So we should check that $2^3$-pole and higher multipole terms less
contribute, though we do not touch this problem in this paper.

\vfill\eject

\section{Discussions}
\par
We have argued the classification of singularities of caustics
based on the multipole expansion model of lensing object.
In the application to the multiple quasar system of PG1115+080
there have arisen several problems.
Fig.8 shows that deamplified images D and E are both normal to
critical line.
Arg$({du \over T_u})$ is given by

\begin{equation}
 \arg({du \over T_u}) = {1 \over 2}\arg(\gamma)
 - \arg( \bar\gamma \partial_{\bar u} \gamma)
\end{equation}

from Eq.(\ref{elongtocrit}) and $\kappa=0$ for multipole expansion.
Whereas amplification is given by

\begin{equation}
 \mu \equiv D^{-1} = {1 \over 1-|\gamma|^2}~.
\end{equation}

Amplification does not depend on the argument of $\gamma$.

\par
Therefore if we are concerned with fold catastrophe we can not relate
the amplification with the image direction relative to the critical line.

\par
We should also be careful to the validity of approximation.
Image D and E are near to the lens position and multipole expansion is
badly convergent.
We pointed out that dipole and quadrupole terms give
the contribution of same order.

\par
It is certainly desirable to consider the same lensing system
based on the several models and to check the consistency of models.

\par
We will compare our results with those of elliptical lens model by
Narasimha et al \cite{NSC}.
Lensing object in \cite{NSC} is modeled as the oblate spheroid

\begin{equation}
 x^2 + y^2 + {z^2 \over {1-e^2}} = a^2~.
\end{equation}

Mass density is given by

\begin{eqnarray}
 \rho(a) &=& \rho_0 (1+{a^2 \over r^2_0})^{-{k \over 2}} ~~~~~
                                        \hbox{for}~~ a / r_c \le n
\nonumber \\
         &=& 0 ~~~~~~~~~~~~~~~~~~~~\hbox{for}~~ a / r_c > n ~,
\end{eqnarray}

where $r_c$ is core size.
We can always set $x$,$y$ coordinates in lens plane and z-axis is
tilted relative to the photon path by an angle $\varphi$.

\par
Then the surface mass density projected on the lens plane is

\begin{equation}
 \Sigma(b) = \sqrt{{{1-e^2} \over {1-e^2 \sin^2 \varphi}}} K_k~,
\end{equation}

where

\begin{equation}
 K_k (b) = \int^{n^2}_{b^2/r^2_c} d\omega
 {\rho_0 r_c \over \sqrt{\omega - ({b \over r_c})^2}}
                            (1+\omega)^{-{k \over 2}}
\end{equation}

and

\begin{equation}
 b^2 = x^2 + {y^2 \over {1-e^2 \sin^2 \varphi}}~.
\end{equation}

In this model the dipole component, $\delta$, vanishes and
we will estimate the quadrupole component, $\xi$, adopted in
\cite{YO} as

\begin{eqnarray}
 \xi &=& {1 \over D^2_{OL}}
 {{\int d\vec x \Sigma(\vec x) z^2} \over
                            {\int d\vec x \Sigma(\vec x)}}  \nonumber \\
 &=& {e^2 \sin^2 \varphi \over 2 D^2_{OL}}
 {  \int^{nr_c}_0 db~b^3 K_k(b) \over
    \int^{nr_c}_0 db~b K_k(b)   }   \nonumber \\
 &\equiv& {{e^2 \sin^2 \varphi} \over 2}{r^2_c \over D^2_{OL}}
                                         {F^{(1)}_k \over F^{(0)}_k}~,
\label{kai}
\end{eqnarray}

where $z$ is given by Eq.(\ref{def}) and where $F^{(l)}_k$ is defined by

\begin{eqnarray}
 F^{(l)}_k  &\equiv&  \int^{n^2}_0 dx~x^l \int^{n^2}_x dy
    {1 \over \sqrt{y-x}}(1+y)^{-{k \over 2}}    \\
 &=&  \int^{n^2}_0 dx~x^l (1+x)^{{1-k \over 2}}
        \int^{\sqrt{{{n^2-x} \over {1+x}}}}_0 du
        (1+u^2)^{-{k \over 2}}~.                  \nonumber
\end{eqnarray}

As is easily checked

\begin{equation}
 F^{(0)}_3 = 2[\ln (\sqrt{n^2+1}+n) - {n \over \sqrt{n^2+1}}]
\end{equation}

and

\begin{equation}
 F^{(0)}_3 + F^{(1)}_3 = {2 \over 3}{n^3 \over \sqrt{n^2+1}}~.
\end{equation}

Therefore we obtain

\begin{equation}
 {F^{(1)}_3 \over F^{(0)}_3} \simeq \left\{
   \begin{array}{ll}
      48.5 & \quad \mbox{for~ n=20} \\
      95.9 & \quad \mbox{for~ n=30}~.
   \end{array} \right.
\end{equation}

$r_c/D_{OL}$ in Eq.(\ref{kai}) is related to
the numerical result by

\begin{equation}
 {r_c \over D_{OL}} = {\theta(\hbox{rad}) \over s}~,
\end{equation}

where $\theta(\hbox{rad})$ is the observed image separation angle
in unit of radian and $s$ is the separation length in unit of core size.

\par
So

\begin{equation}
 \xi = {e^2 \sin^2 \varphi \over 2}({\theta(\hbox{sec}) \over s})^2
          {F^{(1)}_k \over F^{(0)}_k} \times 2.4 \times 10^{-11}
\end{equation}

We adopt the values from Table 1 in \cite{NSC}

\begin{equation}
 k=3~,~~~
 e\sin \varphi = 0.6~,~~~
 n = 20~,~~~
 {4GMD \over r_c^2 c^2} = 32~,~~~\hbox{and}~~~
 s_{BC} = 7.46~,
\end{equation}

where $s_{BC}$ is $s$ between images B and C.
Combining these values with $\theta_{BC} \simeq 1^{\prime \prime}.989$ we
obtain
\begin{equation}
 \xi = 1.51 \times 10^{-11}~.
\end{equation}

If we adopt other values $\xi$ takes the range

\begin{equation}
 \xi_{elliptical} \simeq (2 \sim 4) \times \xi_{multipole}~.
\end{equation}

This seems that quadrupole moments evaluated
by the different models coincide in this order.

\par
We may set $k=1$ or $k=5$ in elliptical lens model.
Especially $k=2$ seems to be important for isothermal model
\cite{KSB}\cite{MT}.
In this case, numerical calculation becomes cumbersome and
will be discussed elsewhere.
Lens models so far considered are somehow empirical.
By applying the several models into the same lensing system,
we can construct the more reliable model of lensing object
and gravitational lensing.

\par
We could not still exhaust the merit of CT in this article.
In the near future, the numbers of observed lensing events will
rapidly increase and we will get more precise information \cite{SDSS}.
CT will play increasingly essential role in that situation.

\vfill\eject

\section*{Acknowledgements}
We have benefited greatly from discussions and communications with
Professors H.Yoshida, M.Omote and K.Tomita and Dr. N.Makino.

\vfill\eject
\begin{center}
{\bf Figure Captions}
\end{center}

\begin{description}
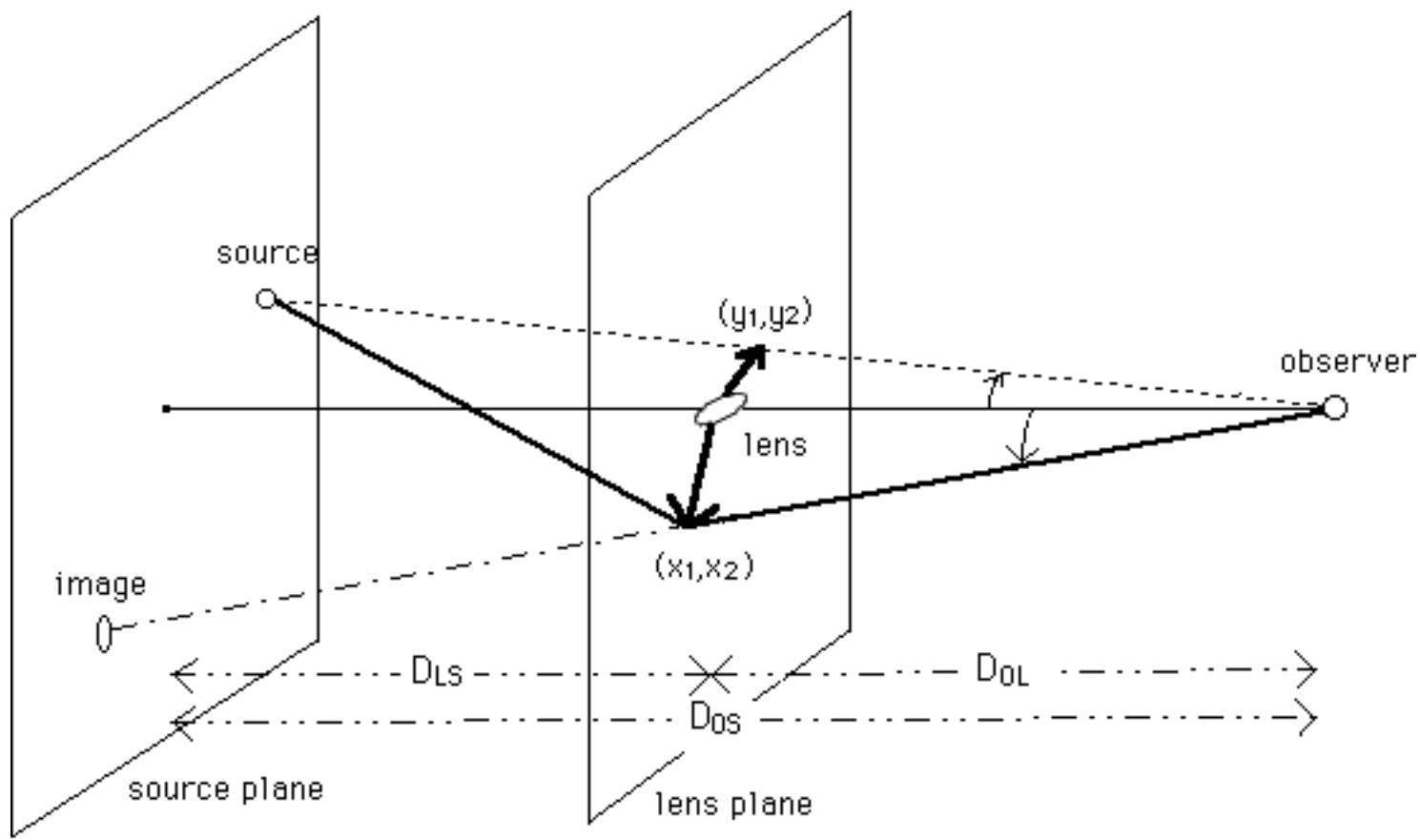
\item[Fig.1] Schematic diagram of the gravitational lens geometry.
             The solid line shows the real path of the light ray.

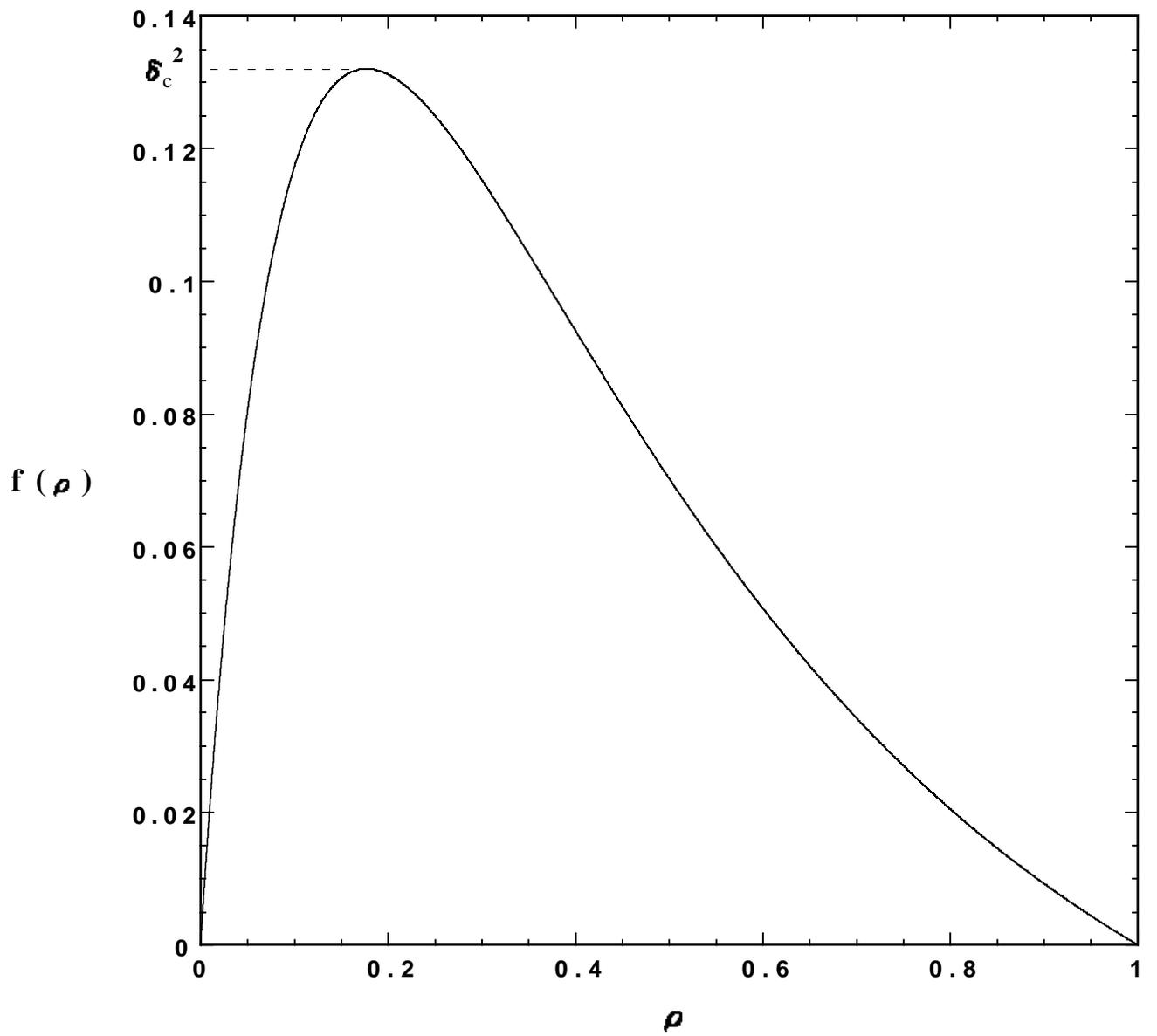
\item[Fig.2] Graph of the function $f(\rho)$ given  in Eq.(\ref{function}).
             Two solutions of $f(\rho)=|\delta|^2$ coalesce at
             $|\delta|^2={10+7\sqrt{7} \over 6^3}$, maximum of $f(\rho)$.

\item[Fig.3] Caustic and critical lines in neighborhood of
$|\delta|=\sqrt{3}/9$ at which beak-to-beak appears on axis.

  \begin{description}
  \item[Fig.3a] $|\delta|=0.17$
  \item[Fig.3b] $|\delta|=\sqrt{3}/9 \sim 0.19$
  \item[Fig.3c] $|\delta|=0.21$
  \end{description}

\item[Fig.4] Caustic and critical lines in neighborhood of
$|\delta|=\delta_c={\sqrt{10+7\sqrt{7}} \over 6\sqrt{6}}$
at which swallowtail appears.
Points P in Fig.4a are the points of cusp singularities which change to
swallowtail when $|\delta|$ increases.
  \begin{description}
  \item[Fig.4a] $|\delta|=0.32$  
  \item[Fig.4b] $|\delta|=\delta_c \sim 0.36$
  \item[Fig.4c] $|\delta|=0.50$
  \end{description}

\item[Fig.5] Caustic and critical lines in neighborhood of
$|Q|=1/24$ at which beak-to-beak appears on $Y$-axis.
  \begin{description}
  \item[Fig.5a] $|Q|=0.03$
  \item[Fig.5b] $|Q|=1/24 \sim 0.04$
  \item[Fig.5c] $|Q|=0.05$
  \end{description}

\item[Fig.6] Caustic and critical lines in neighborhood of
$|Q|=Q_1=1/8$ at which butterfly appears.
Point P in Fig.6b is the point where butterfly appears.
  \begin{description}
  \item[Fig.6a] $|Q|=0.12$
  \item[Fig.6b] $|Q|=1/8$
  \item[Fig.6c] $|Q|=0.13$
  \end{description}

\item[Fig.7] Caustic and critical lines in neighborhood of
$|Q|=Q_2={\sqrt{598+82\sqrt{41}}\over 192}$
at which swallowtail appears.
  \begin{description}
  \item[Fig.7a] $|Q|=0.17$
  \item[Fig.7b] $|Q|=Q_2 \sim 0.1745$
  \item[Fig.7c] $|Q|=0.18$
  \end{description}

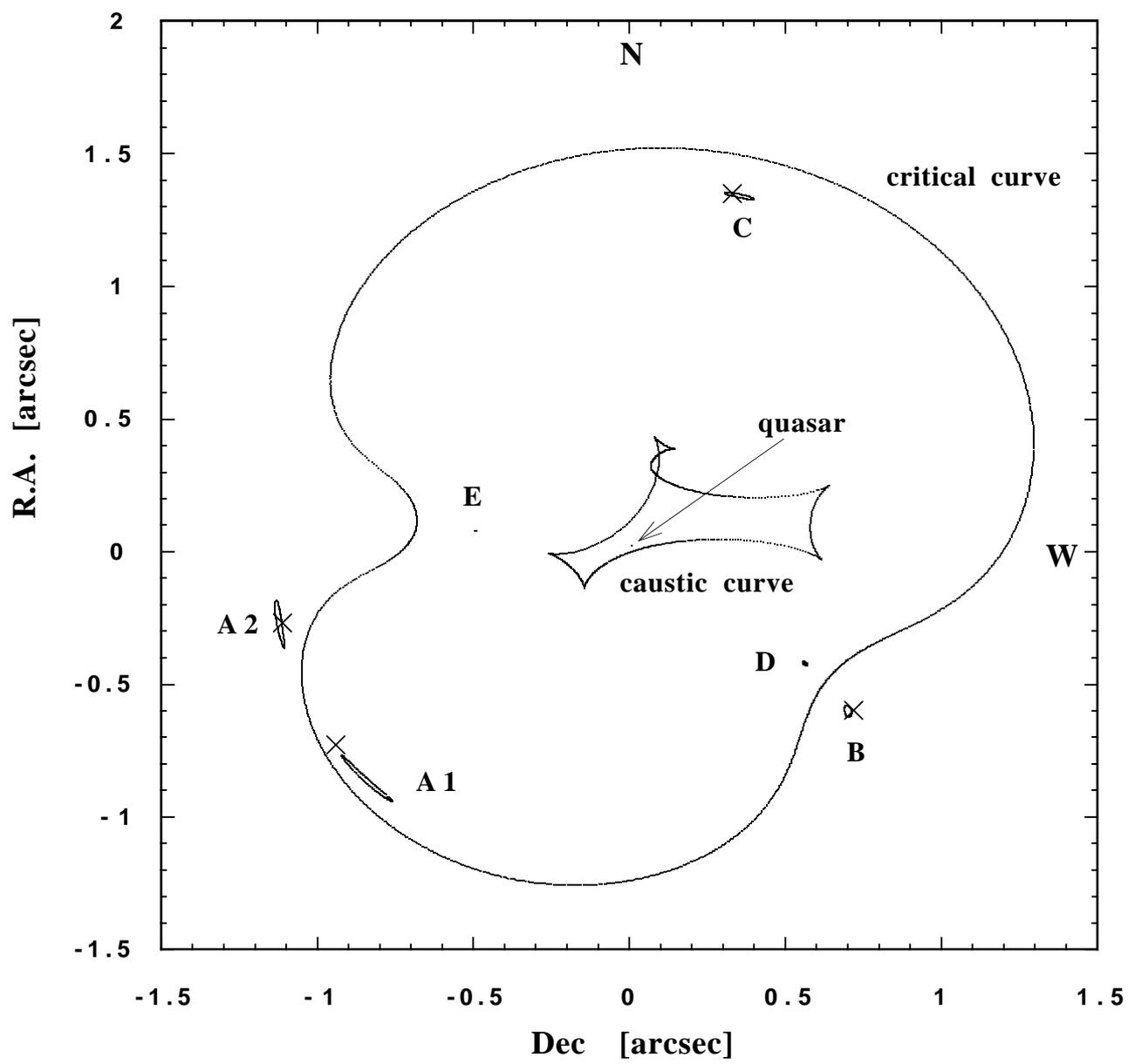
\item[Fig.8] Shapes of the images, critical and caustic line
 for the multiple quasar
PG1115+080 from our calculation. The lensing galaxy is at
coordinate origin.
Observed image positions are denoted by crosses.

\item[Table 1] Observed image positions and their amplifications of PG1115+080.

\end{description}

\vfill\eject\
\thispagestyle{empty}

\begin{table}
\begin{center}
  \begin{tabular}{|c|r@{.}l|r@{.}l|r@{.}l|} \hline
   image &
  \multicolumn{2}{|c|}{$x_1$(sec)} &
  \multicolumn{2}{|c|}{$x_2$(sec)} &
  \multicolumn{2}{|c|}{$\mu / \mu_C$} \\ \hline
  $A_1$ & $-1^{\prime \prime}$ & $27$ & $-2^{\prime \prime}$
               & $08$ & $3$ & $22$ \\ \hline
  $A_2$ & $-1^{\prime \prime}$ & $44$ & $-1^{\prime \prime}$
               & $62$ & $2$ & $49$ \\ \hline
  $B$   & $0^{\prime \prime}$ & $39$ & $-1^{\prime \prime}$
               & $95$ & $0$ & $64$ \\ \hline
  $C$   & $ 0^{\prime \prime}$ & $00$ &  $0^{\prime \prime}$
               & $00$ & $1$ & $00$ \\ \hline\hline
  $\hbox{lens}$   & $-0^{\prime \prime}$ & $33$ & $-1^{\prime \prime}$
               & $35$ &
           \multicolumn{2}{|c|}{} \\ \hline
  \end{tabular}
\end{center}
\caption{Observational data}
\label{tab1}
\end{table}

\vfill\eject
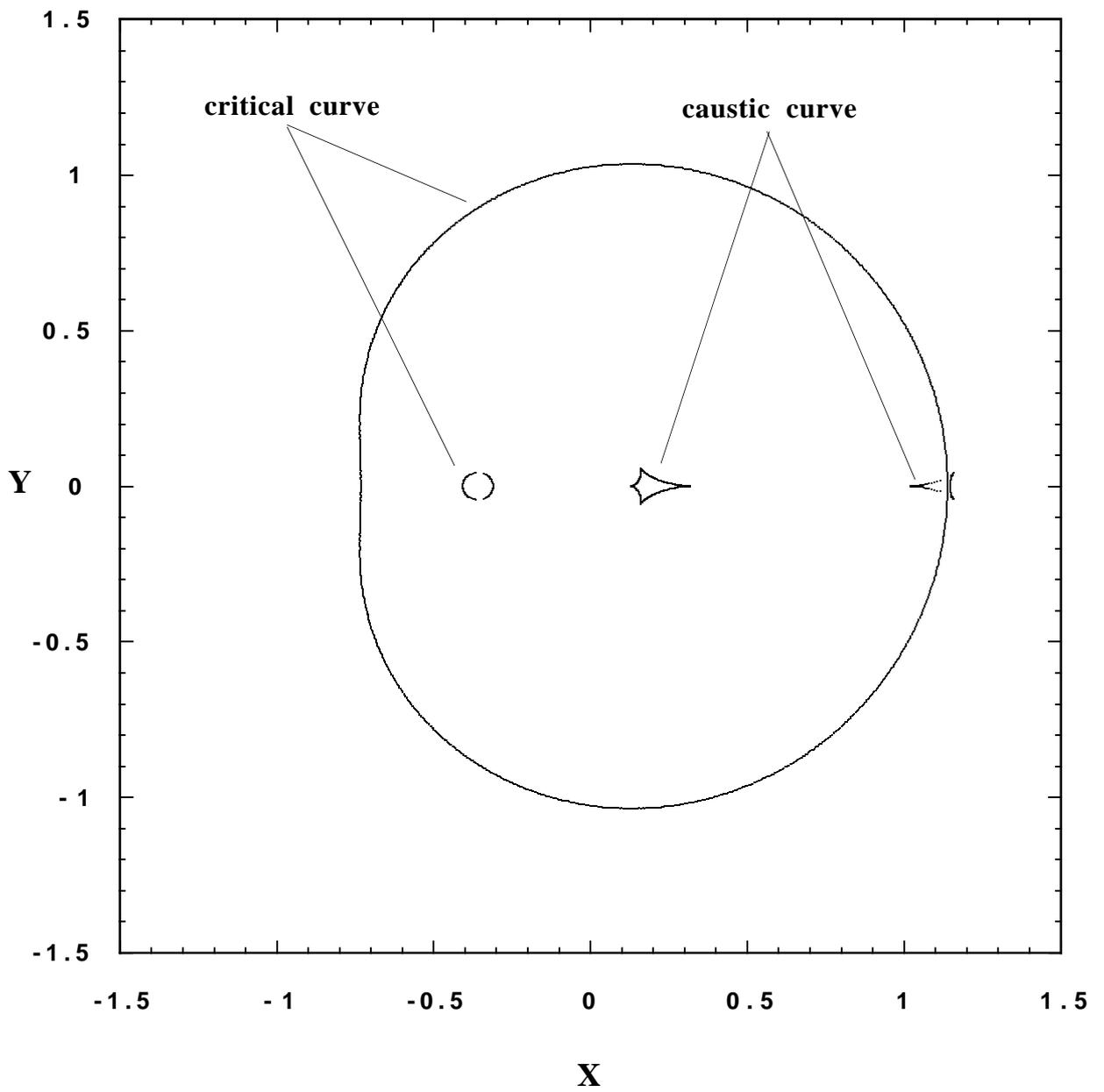
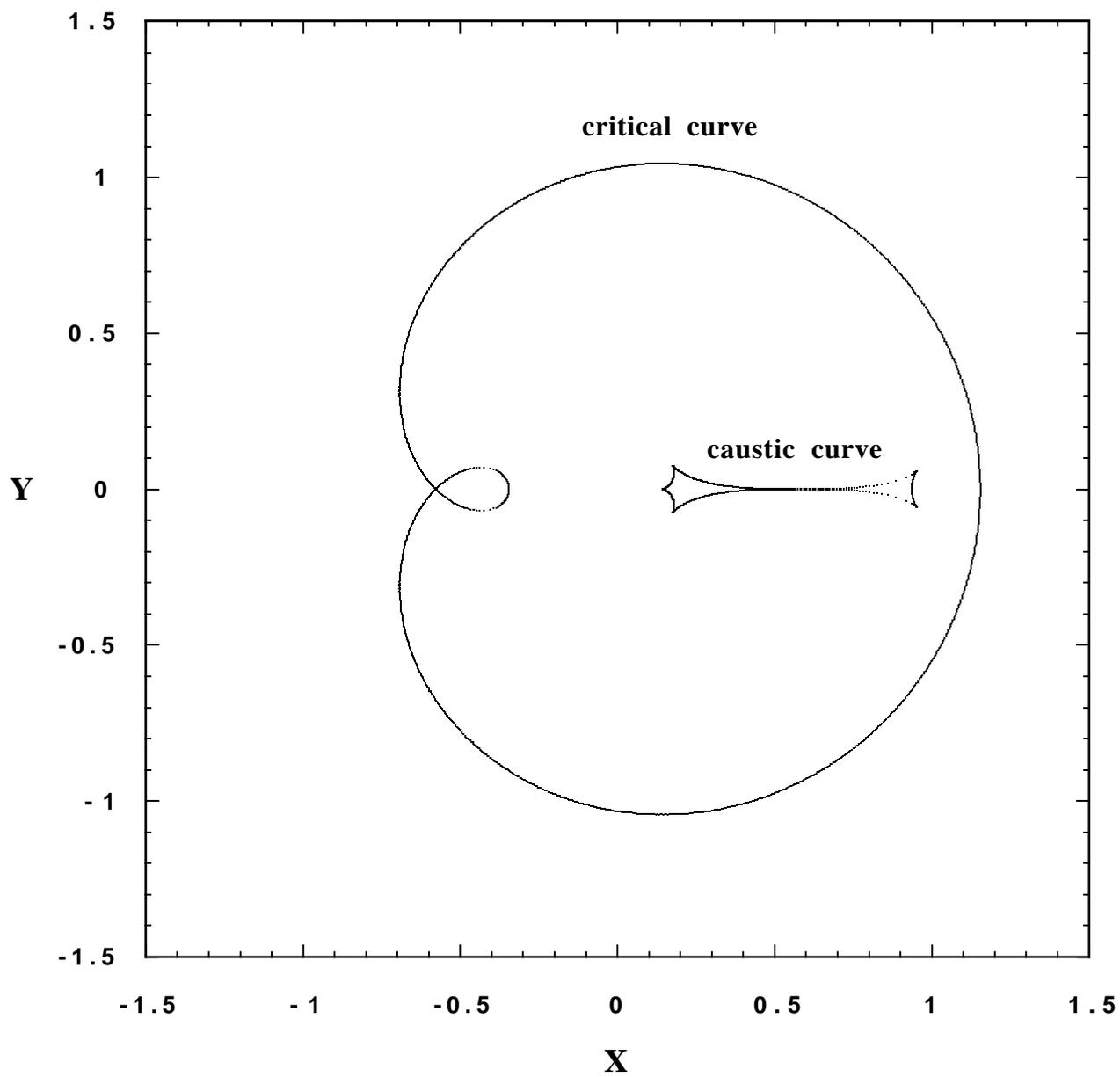
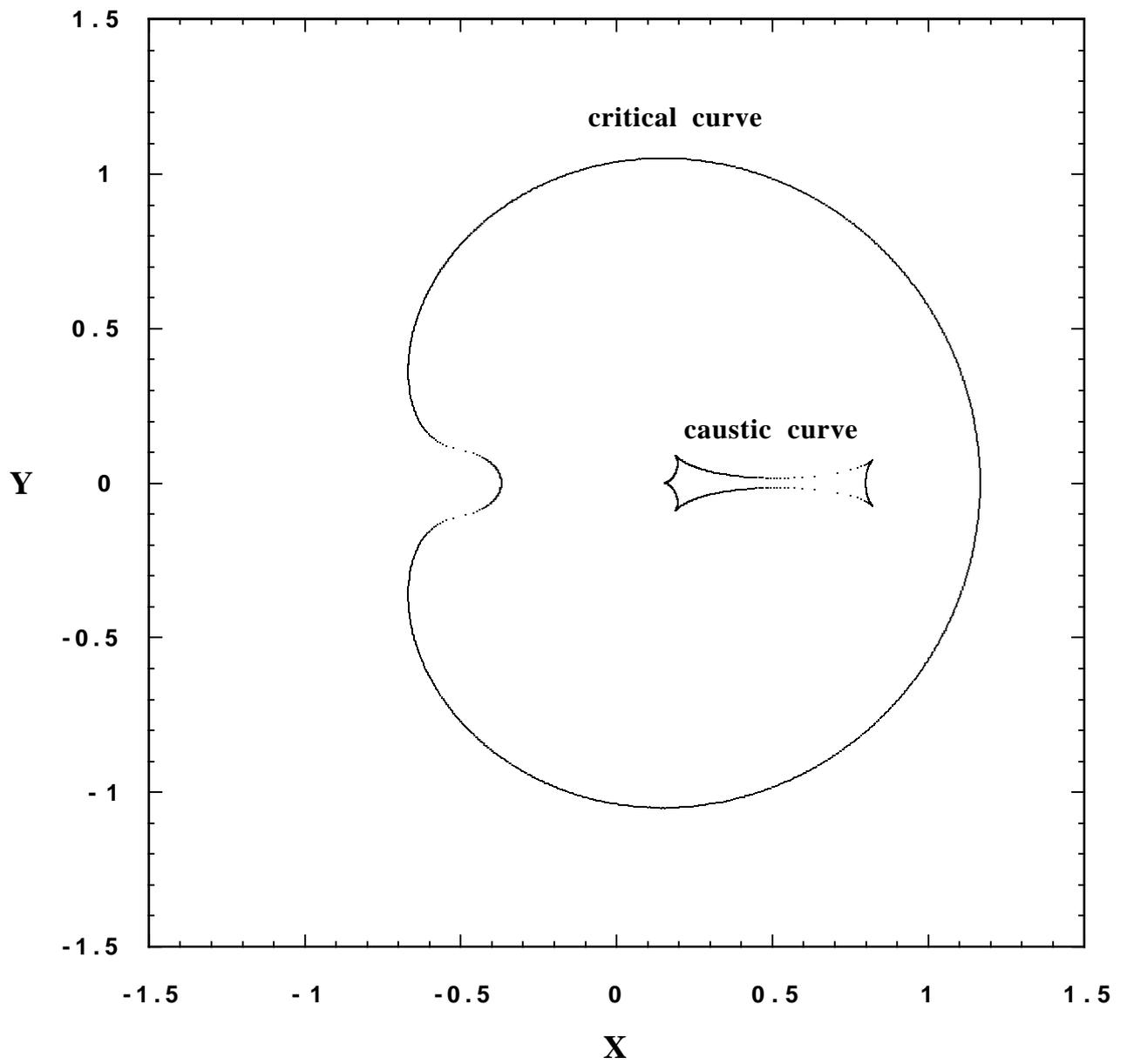
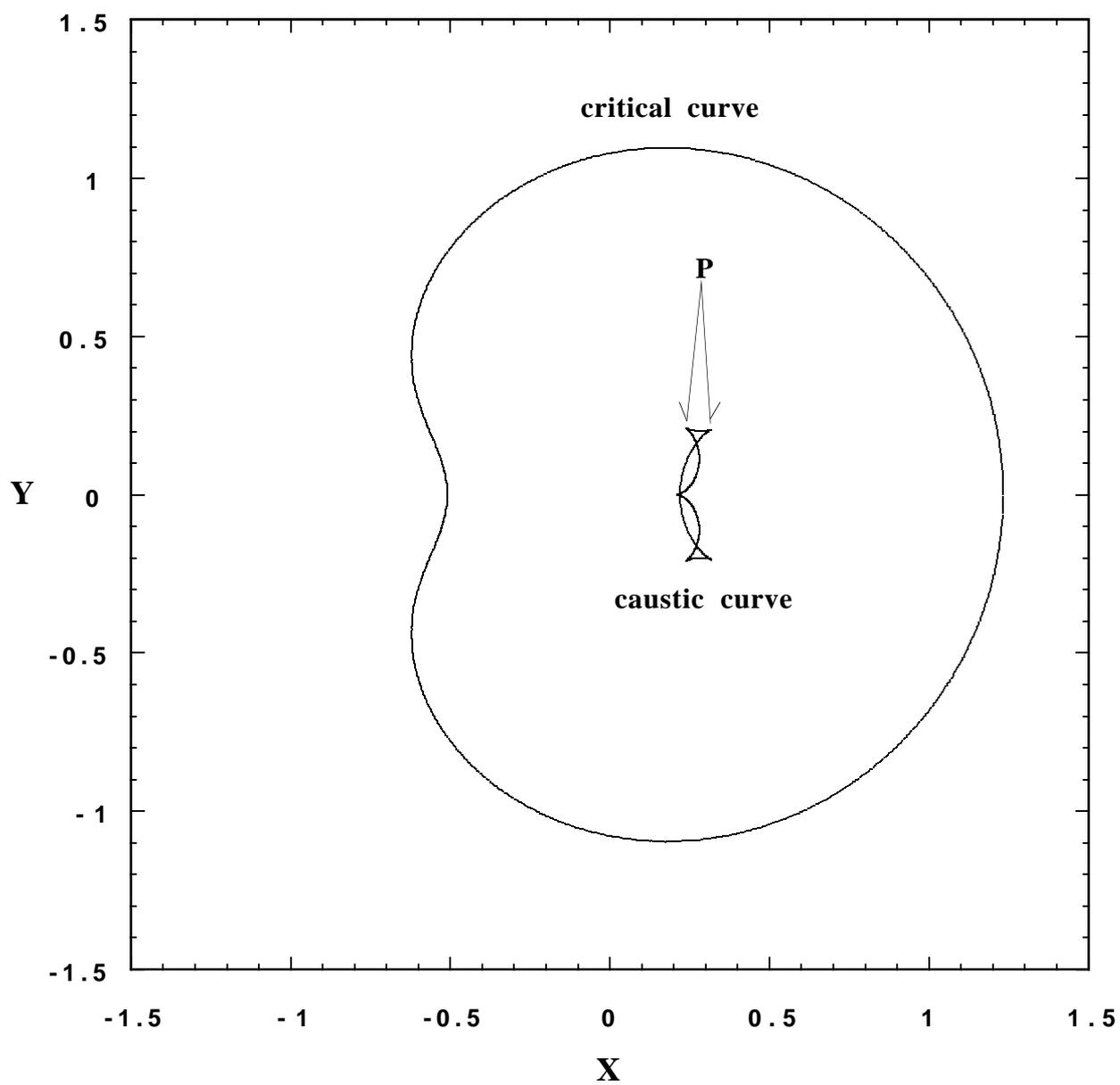
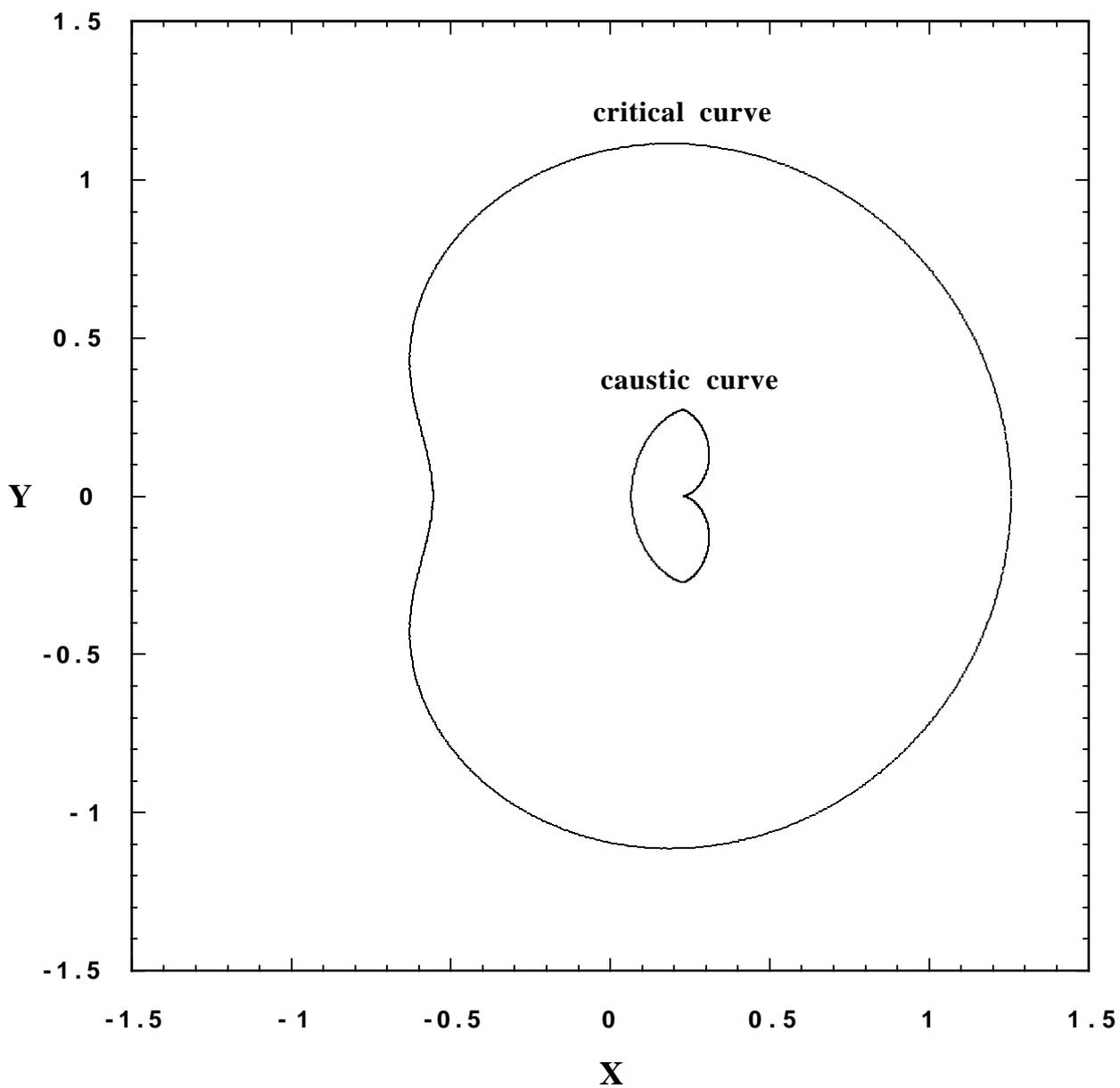
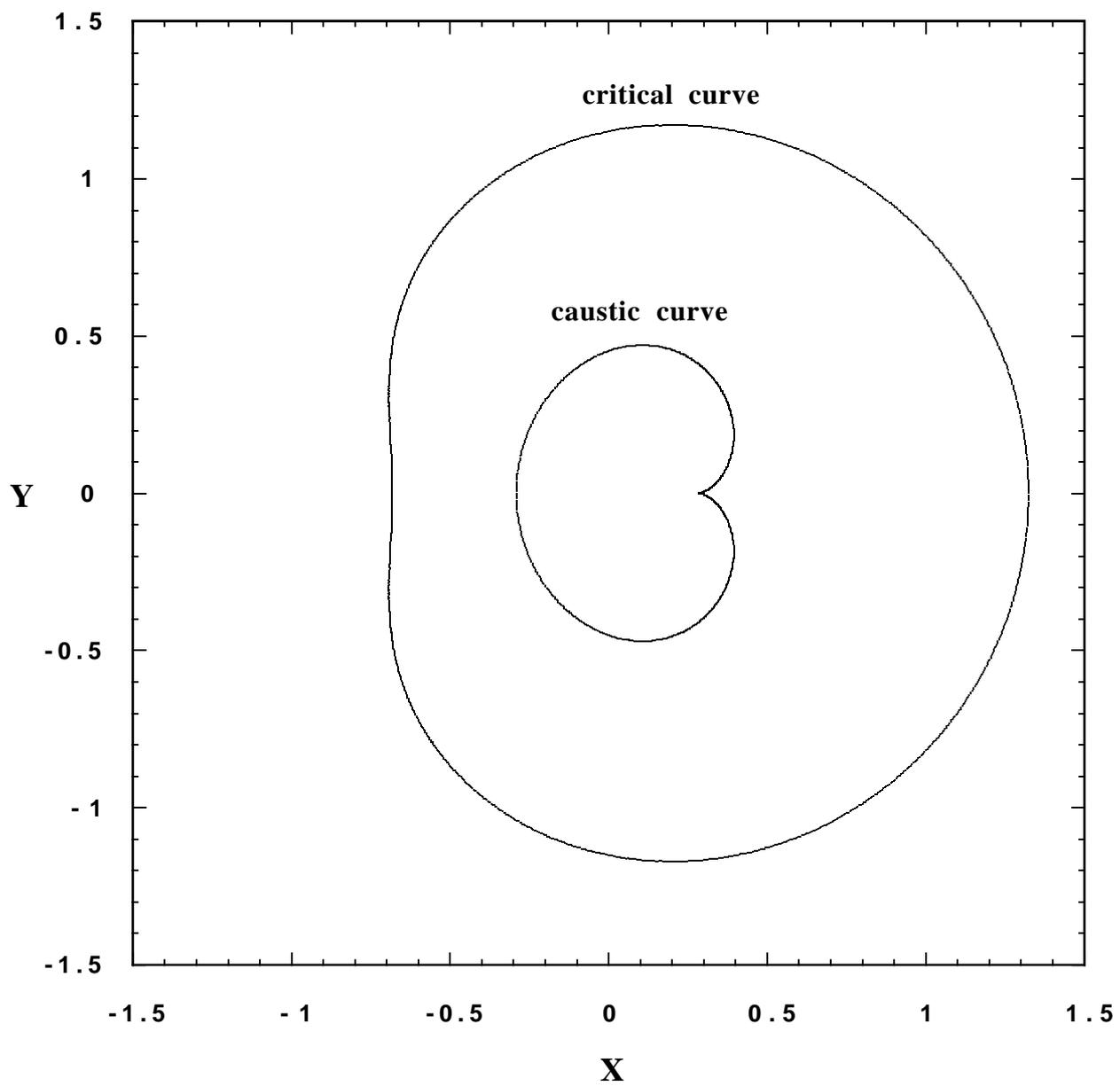
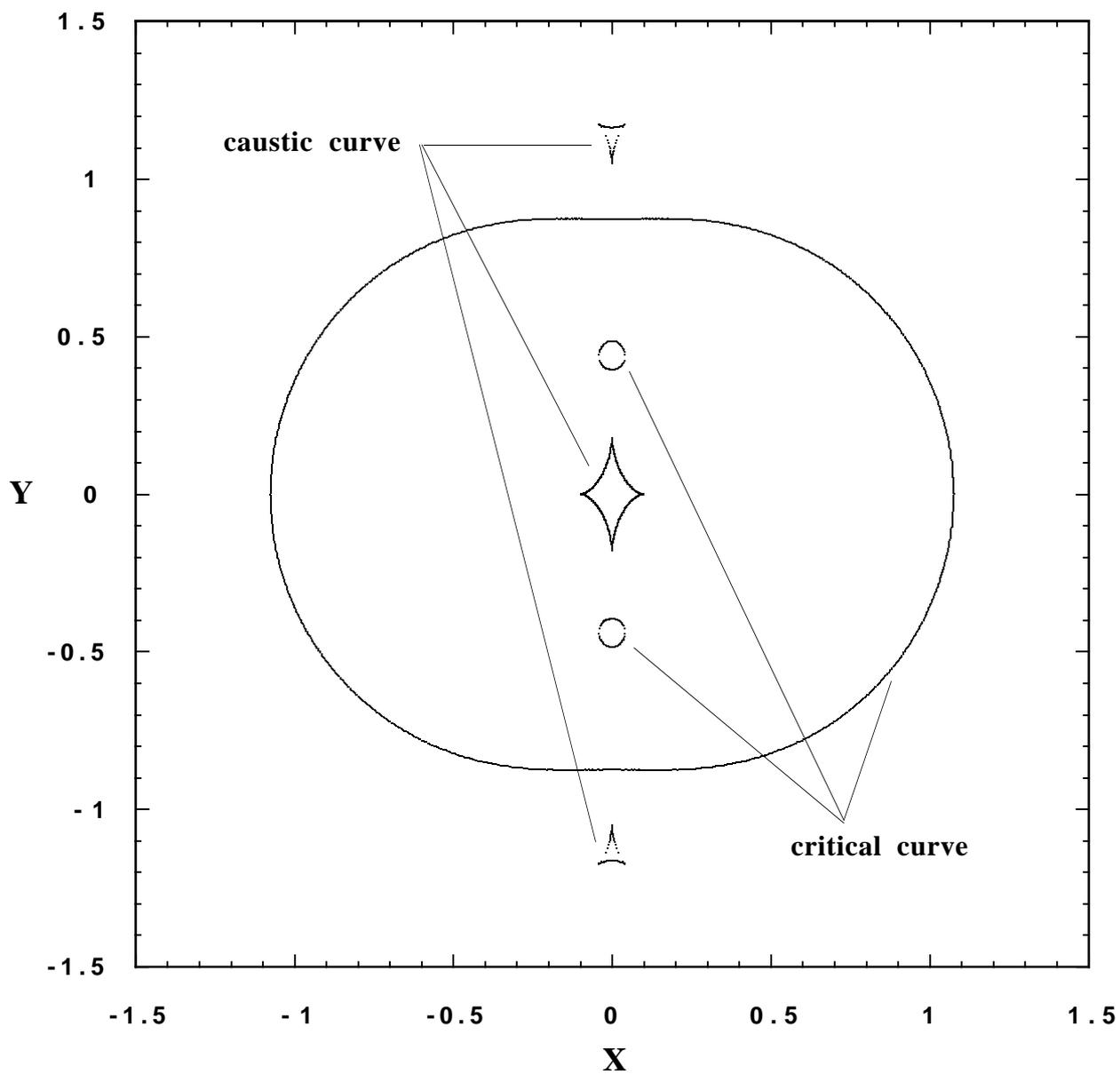
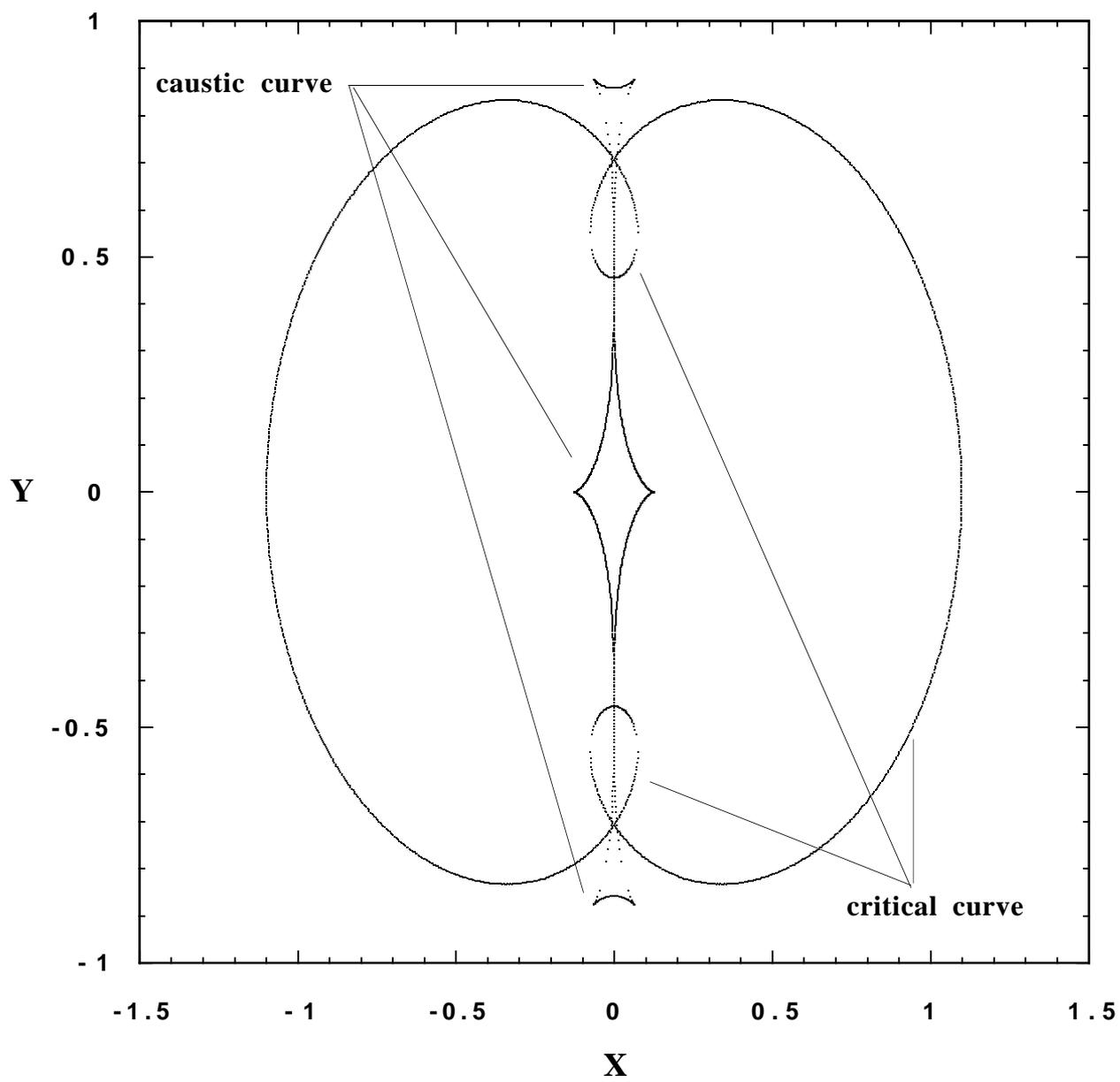
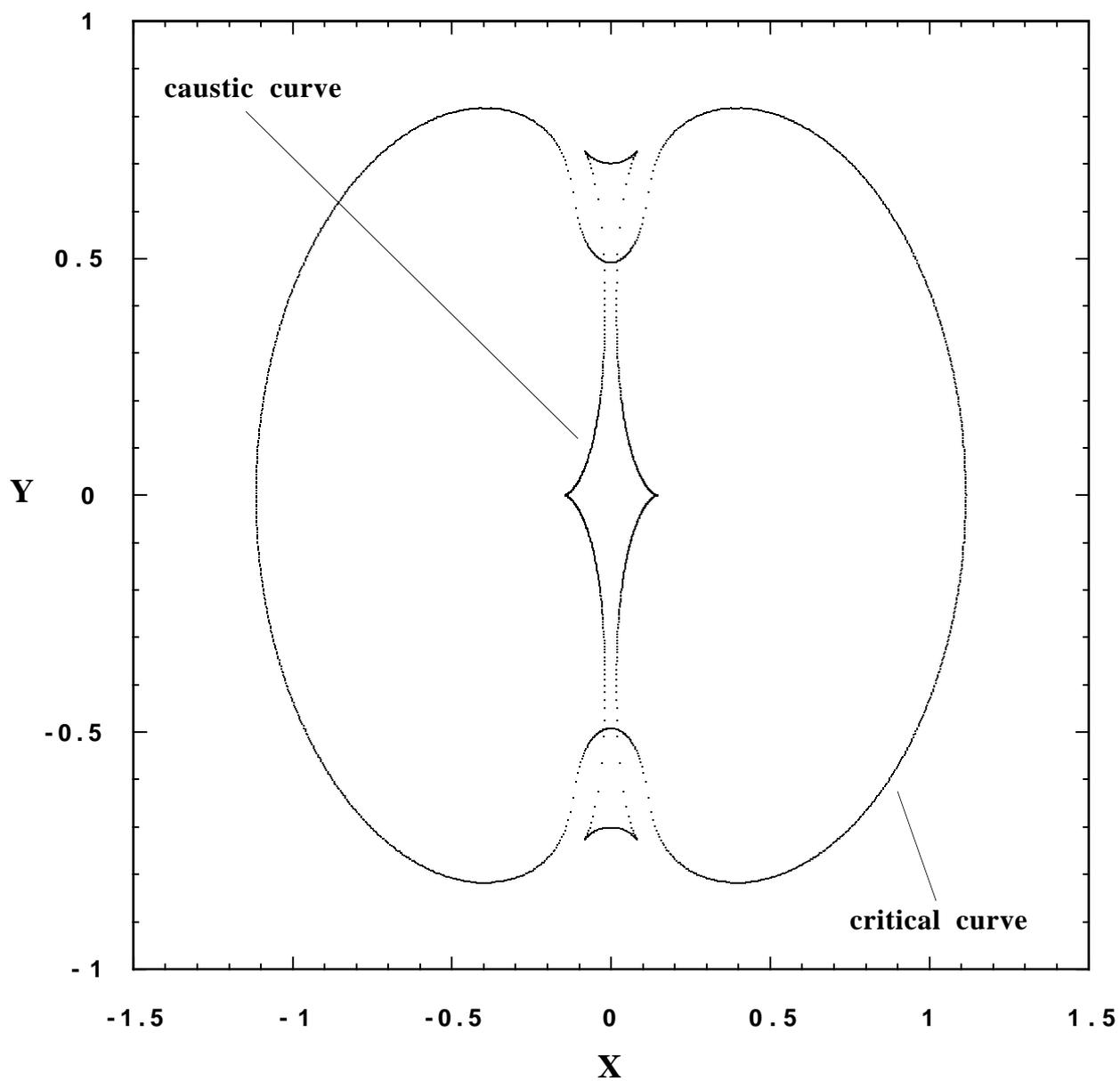
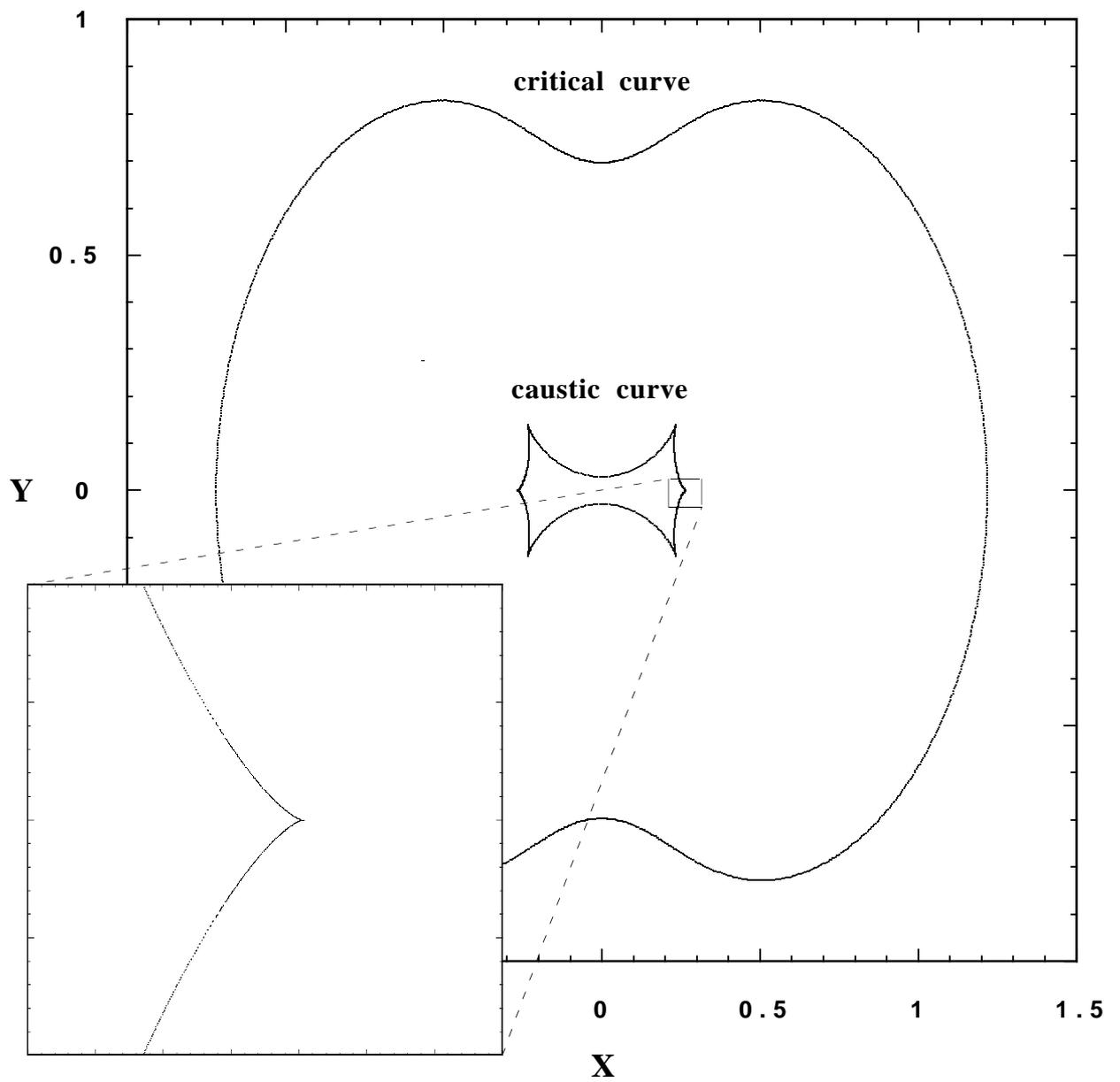
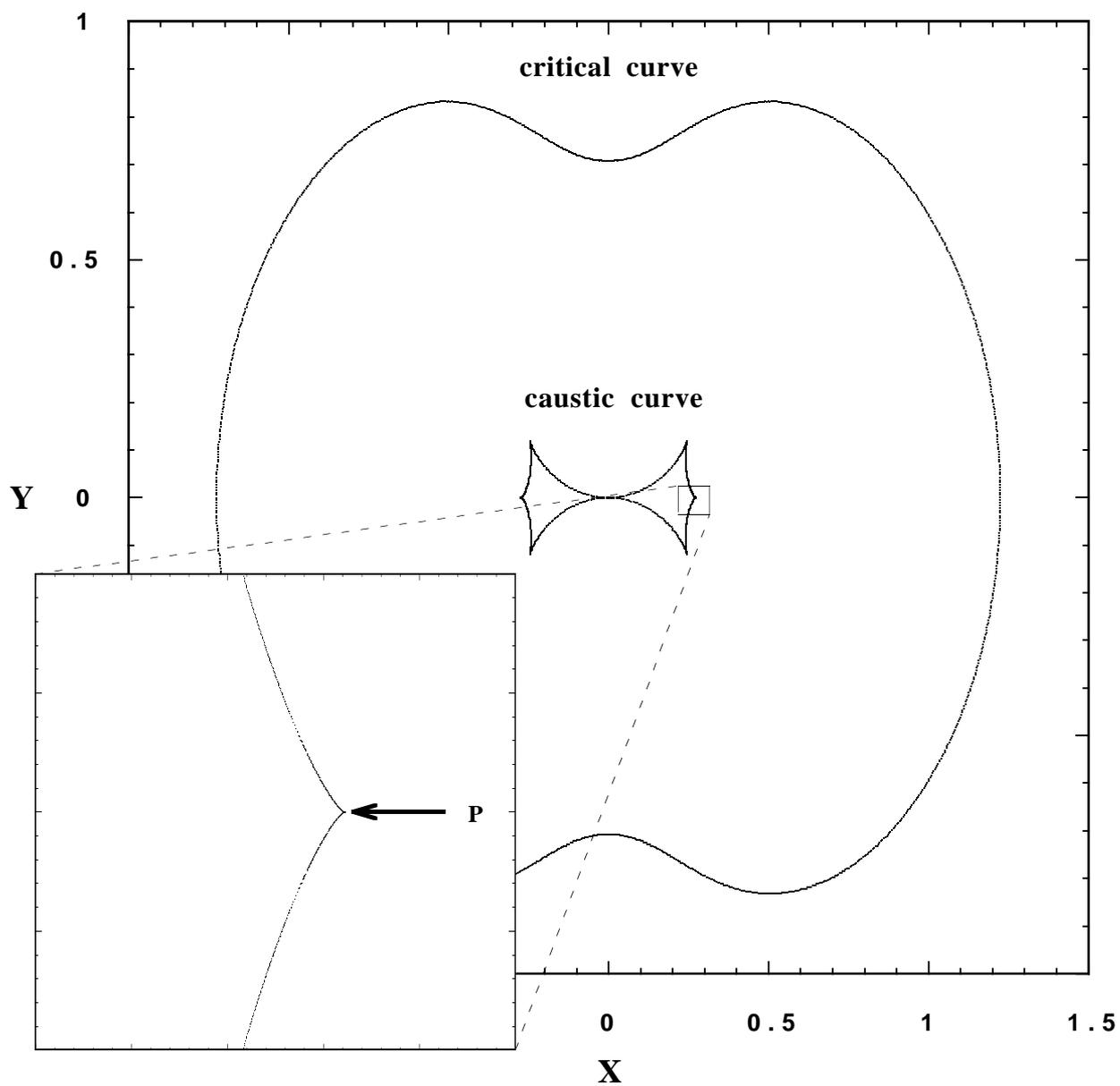
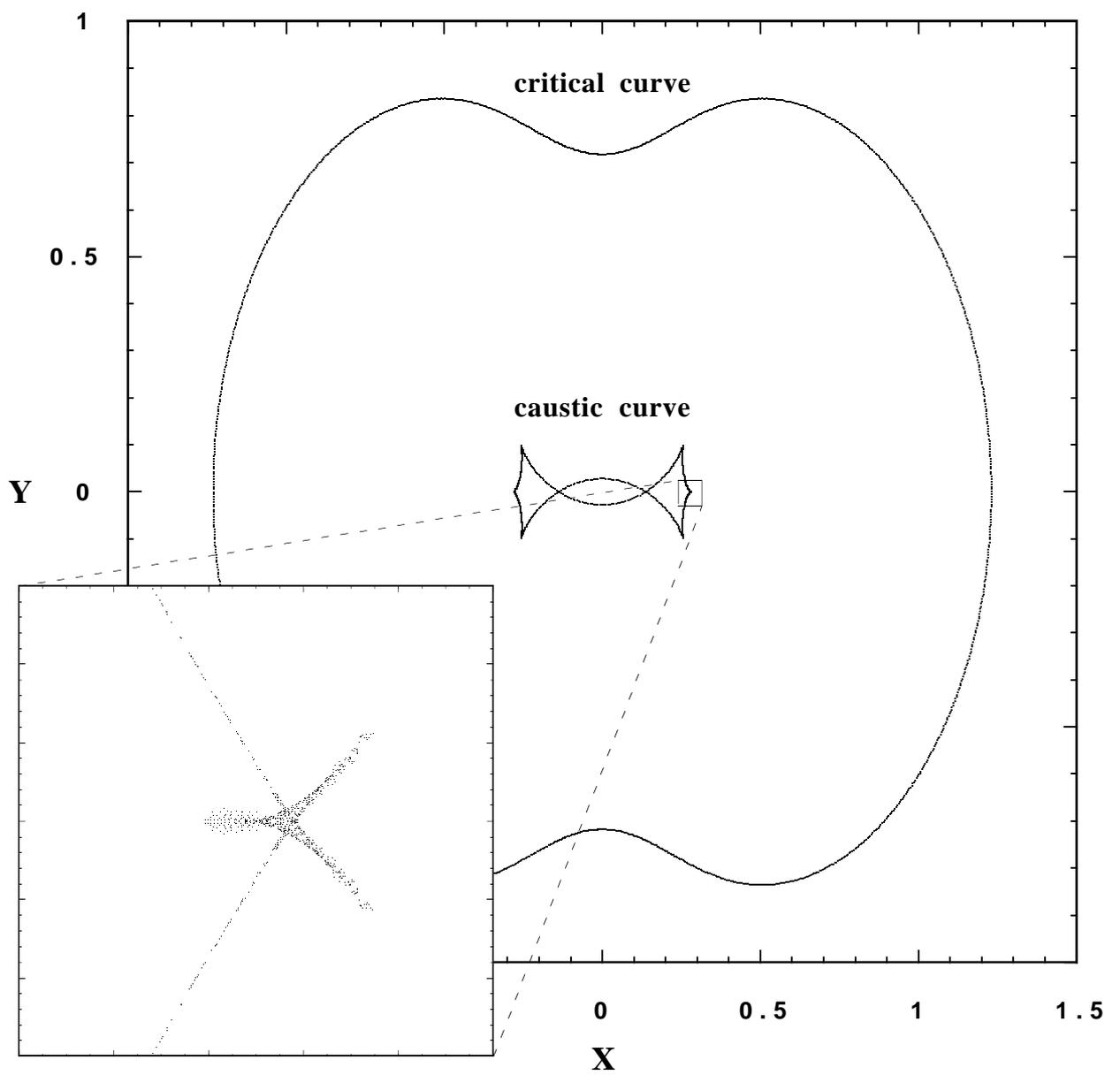
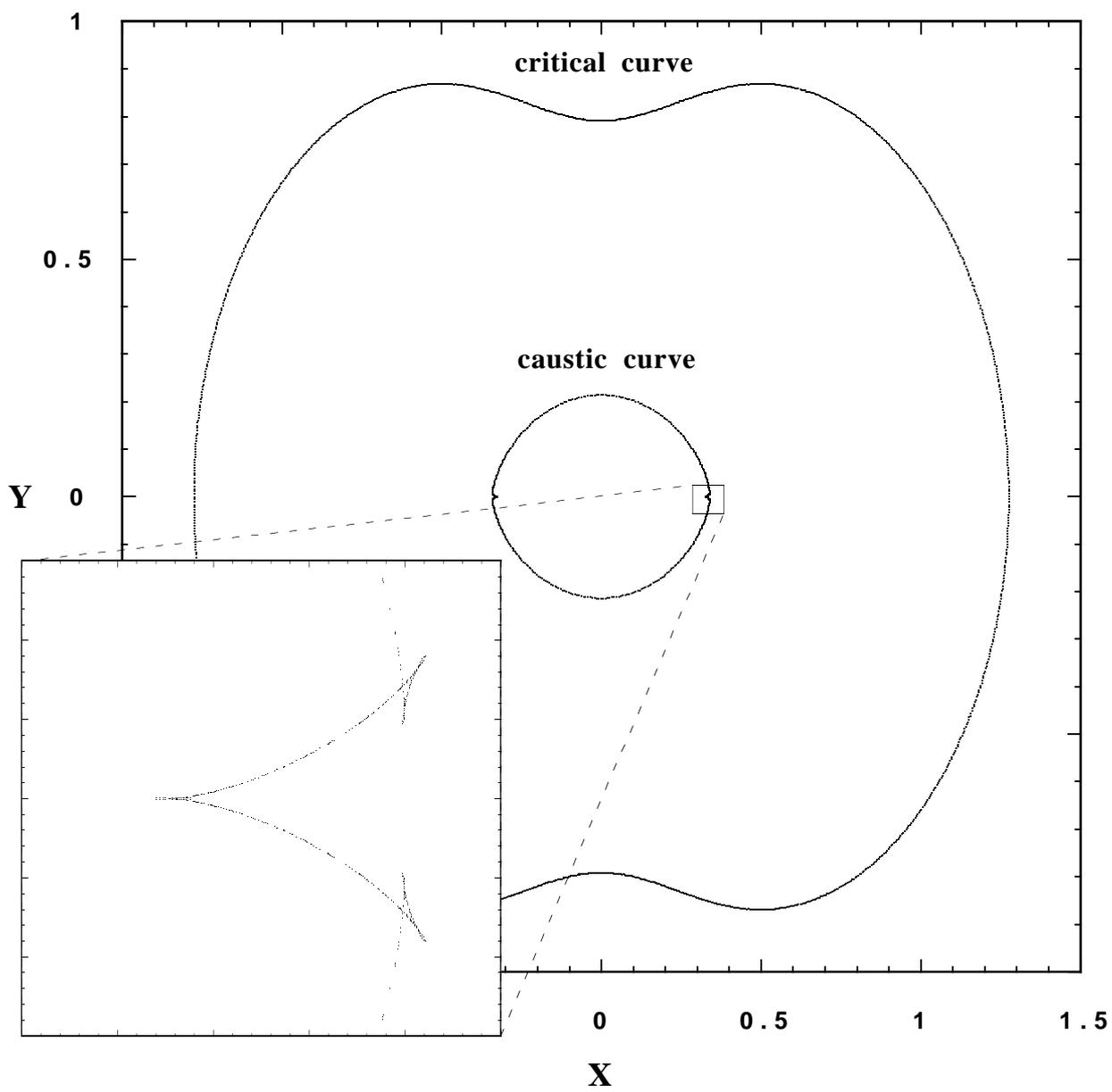
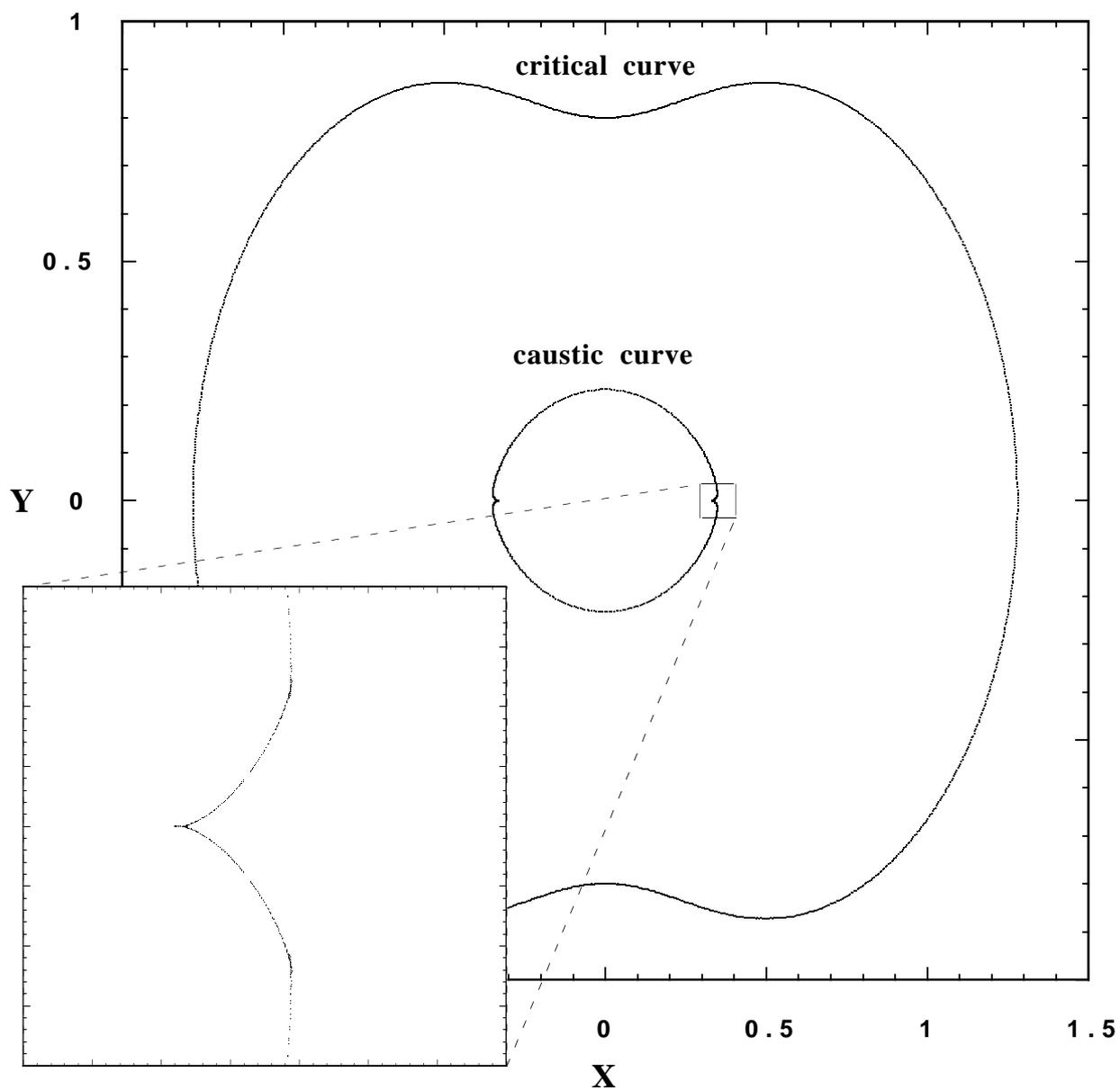
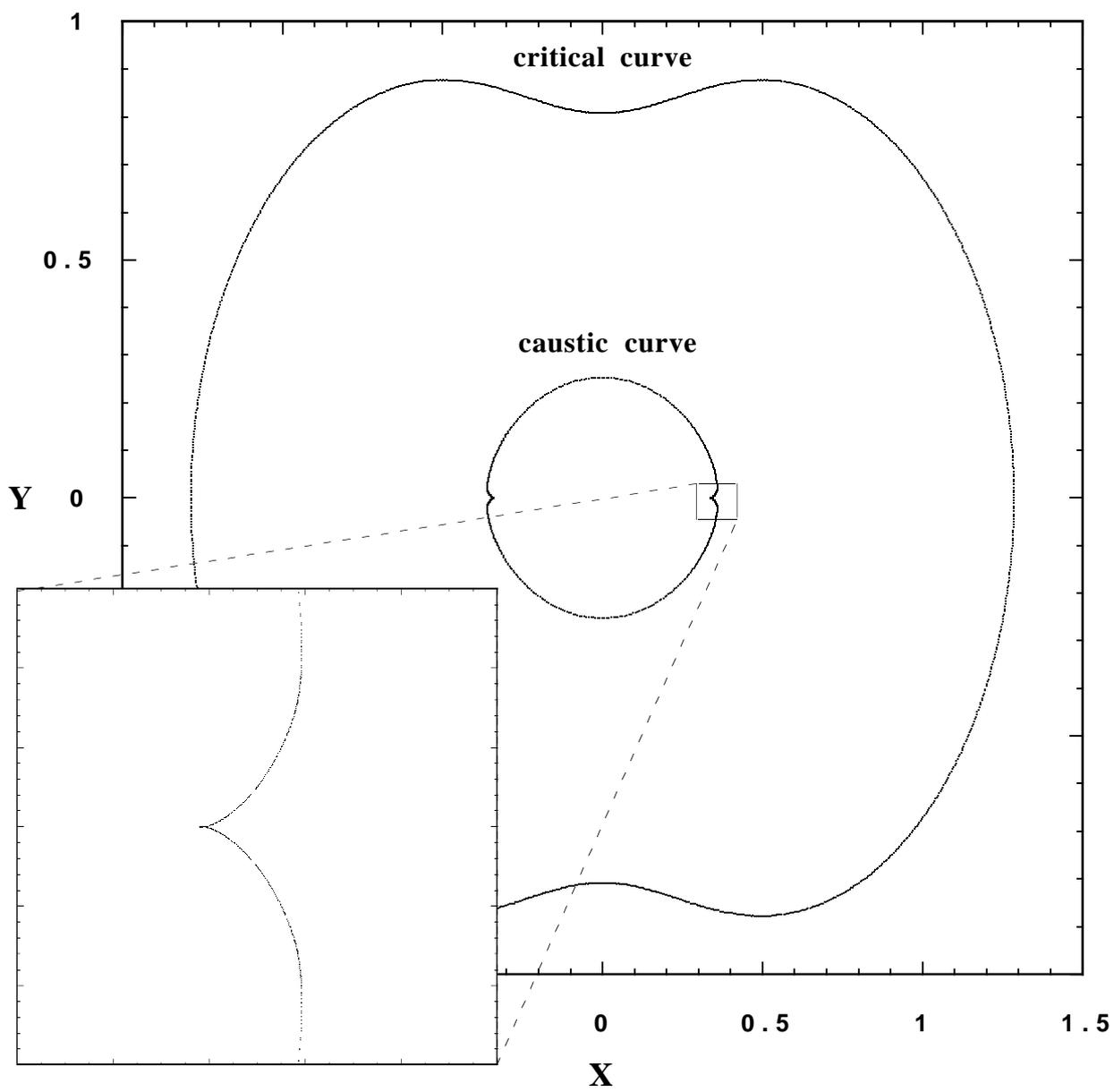
\end{document}